\documentclass[prd,aps,eqsecnum,amsmath,floatfix,nofootinbib,preprint,tightenlines]{revtex4-2}

\usepackage{latexsym}
\usepackage{graphicx}
\usepackage[dvipsnames]{xcolor}

\def\mynext{\smallskip\noindent }

\usepackage{hyperref}

\def\floatcaption#1#2{ \caption{#2 \label{#1}} }

\def\bibi{\bibitem}

\def\ttl#1{{\it #1}}

\def\c{\chi}
\def\d{\delta}
\def\e{\epsilon}                
\def\f{\phi}                    
\def\g{\gamma}

\def\j{\psi}

\def\l{\lambda}
\def\m{\mu}
\def\n{\nu}
\def\o{\omega}
\def\p{\pi}                     
\def\th{\theta}                  
\def\s{\sigma}                  
\def\t{\tau}

\def\x{\xi}

\def\J{\Psi}

\def\O{\Omega}

\def\U{\Upsilon}

\def\ca{{\cal A}}
\def\cb{{\cal B}}

\def\cd{{\cal D}}
\def\ce{{\cal E}}

\def\ch{{\cal H}}   

\def\cn{{\cal N}}
\def\co{{\cal O}}

\def\car{{\cal R}}

\def\cu{{\cal U}}

\def\bo{\raisebox{-.4ex}{\large$\Box$}}                 
\def\cbo{{\,\raise-.15ex\Sc [\,}}                       

\def\sl#1{\rlap{\hbox{$\mskip 1 mu /$}}#1}      

\def\sbra#1{\left\langle #1\right|}             
\def\sket#1{\left| #1\right\rangle}             
\def\svev#1{\left\langle #1\right\rangle}       

\def\ddt#1{{\buildrel {\hbox{\LARGE .\kern-2pt.}} \over {#1}}}

\def\ie{\mbox{\it i.e.}}
\def\eg{\mbox{\it e.g.}}

\def\hc{{\rm h.c.\,}}

\def\half{{1\over 2}}
\def\Re{{\rm Re\,}}
\def\Im{{\rm Im\,}}

\def\Fig#1{Fig.~\ref{#1}}

\def\bj{\overline\psi}

\def\bx{\overline\x}

\def\hH{\hat{H}}

\def\tP{\tilde{P}}

\def\tq{\tilde{q}}
\def\tl{\tilde\ell}
\def\ts{\tilde\s}
\def\dirac{\eta}
\def\bdirac{\overline\eta}

\def\textit#1{{\it \!\!\! #1 \!\!}}

\def\U{{\rm U}}

\def\Heff{H_{\rm eff}}

\def\Hint{H_{\rm int}}
\def\Oirr{O_{\rm irr}}
\def\ZZWY{ZZWY}
\def\myvec#1{\vec{#1}}
\def\taxi#1{\| #1 \|}
\def\Emin{E_{\rm min}}

\begin{document}

\begin{center}
\vspace{10mm}
{\large\bf Constraints on the symmetric mass generation paradigm\\[2mm]
            for lattice chiral gauge theories }\\[8mm]
Maarten Golterman$^a$ and Yigal Shamir$^b$\\[8 mm]
$^a$Department of Physics and Astronomy, San Francisco State University,\\
San Francisco, CA 94132, USA\\
$^b$Raymond and Beverly Sackler School of Physics and Astronomy,\\
Tel~Aviv University, 69978, Tel~Aviv, Israel\\[10mm]
\end{center}

\begin{quotation}
Within the symmetric mass generation (SMG) approach to the construction
of lattice chiral gauge theories, one attempts to use interactions to render
mirror fermions massive without symmetry breaking, thus obtaining the desired
chiral massless spectrum.  If successful,
the gauge field can be turned on, and thus
a chiral gauge theory can be constructed in the phase in which SMG takes place.
In this paper we argue that the zeros that often replace
the mirror poles of fermion two-point functions in an SMG phase should be
``kinematical'' singularities.  We conjecture that the SMG interactions
generate opposite-chirality bound states, which combine
with the gapped elementary mirror states to form massive Dirac fermions.
The propagator zeros can then be avoided by choosing an
appropriate set of interpolating fields that contains both elementary
and composite fields.  This allows us to apply general constraints
on the existence of a chiral fermion spectrum which are valid
in the presence of (non-gauge) interactions of arbitrary strength,
including in any SMG phase.
Using a suitably constructed one-particle lattice hamiltonian describing the
fermion spectrum, we formulate a generalized
no-go theorem which establishes the conditions for
the applicability of the Nielsen-Ninomiya theorem to this hamiltonian.
If these conditions are satisfied, the massless fermion spectrum
must be vector-like.  We add some general observations
on the strong coupling limit of SMG models.  We also elaborate
on the qualitative differences between four-dimensional and
two-dimensional theories that limit the lessons that can be drawn
from two-dimensional models.  Finally, we compile a list of open questions
which must be addressed in any SMG model in order to determine
whether or not it is subject to the generalized no-go theorem.
\end{quotation}

\newpage
\section{\label{intro} Introduction}
Because of the fermion species-doubling problem,
the nonperturbative construction of chiral gauge theories
on the lattice is a long-standing challenge.\footnote{%
  For reviews, see Refs.~\cite{YSrev,MGrev,mirrorrev}.
}
The physical origin of species doubling
was first addressed by Karsten and Smit \cite{KS},
tying the phenomenon to the chiral anomaly,
and then generalized by Nielsen and Ninomiya \cite{NN}.
Since then, the program of putting chiral gauge theories on the lattice
has gained partial successes, but no completely worked-out
method for doing so exists at present.

Building on earlier work by Ginsparg and Wilson \cite{GW},
by Kaplan \cite{DWF}, and by Narayanan and Neuberger
\cite{NNchiralPRL,NNchiralNPB,HNoverlap,HNGW}, L\"uscher
successfully constructed anomaly-free abelian chiral gauge theories, while
requiring one new algebraic constraint on the fermion spectrum beyond
the familiar anomaly-cancellation condition \cite{MLabelian}.
As for nonabelian chiral gauge theories, he was able to define them
to all orders in lattice perturbation theory \cite{MLnonabelian,MLpertth}.

A second approach is being pursued by Kaplan and collaborators.
In this approach, the chiral fermions reside on a four-dimensional boundary
of a five-dimensional space, with massive fermion degrees of freedom
inside the five-dimensional bulk.  The dynamical gauge field
lives on the same boundary as the chiral fermions, and is extended
into the five-dimensional bulk via a classical differential equation, for which
the dynamical four-dimensional gauge field provides the boundary values.
The goal is to dampen the gauge field inside the five-dimensional bulk
such that, when the fermion spectrum is anomaly free,
the long-distance physics would originate exclusively from the coupling
of the dynamical gauge field to the chiral fermions at the boundary.
Concrete realizations of this approach
were proposed in a domain-wall fermion setup \cite{GK1,GK2},
and more recently in a disk setup \cite{Kd,KSr1}.  The scope of this approach
is presently under investigation \cite{GScrnt,KSr2}.

A third approach, which we have pursued, is the gauge-fixing approach.
The chiral gauge invariance is explicitly broken on the lattice.
Gauge invariance and (conjecturally) unitarity are restored only
in the continuum limit, provided that the fermion
spectrum is anomaly free.  The inclusion of a suitable gauge-fixing
lattice action, as well as, in the nonabelian case, ghost fields, ensures
the existence of a novel critical point where the target chiral gauge theory
can be defined \cite{YS95,GFaction,BGSphase,BGSPRL,BGSlat97b,FNV,nachgt}.
The current challenges of this approach have to do primarily
with the dynamics of the gauge-fixing sector in the nonabelian case
\cite{gtilde,dimtrans}.

The fourth, and historically the first, approach has a long history.
It has evolved into what nowadays is usually called the
symmetric mass generation (SMG) approach.\footnote{%
  For a review of the more recent work on SMG, see Ref.~\cite{WYreview}.
}
One starts from a lattice gauge theory with massless Dirac fermions.
The (anomaly free) fermion spectrum of the target chiral gauge theory would be
obtained by selectively retaining only one of the two chiralities
of each Dirac fermion; this is the physical fermion, while the unwanted
opposite-chirality component is the ``doubler,'' or ``mirror'' fermion.
In order to recover the target chiral gauge theory
in the continuum limit one must therefore find a way to decouple
all the mirror fermions.
The SMG paradigm envisages that this goal can be achieved by adding to
the lattice action judicially chosen strong interactions that do not involve
the gauge field.  They can be multi-fermion interactions and/or depend on
additional scalar fields introduced especially for this purpose.

Let the gauge symmetry be a compact Lie group $G$.
If we turn off the gauge field, we obtain a so-called {\em reduced model}
in which $G$ is an exact global symmetry.
The fermion content of the original gauge theory can be read off by assigning
the fermion states of the reduced model to representations of $G$,
if $G$ is nonabelian; or by specifying the (integer valued)
charge of each fermion state, if $G$ is abelian.  The reduced model
will typically have a nontrivial phase diagram, and different phases
may in principle have a different massless fermion spectrum.
The goal of the SMG framework is thus to achieve a novel ``SMG phase''
with the following features:
the global $G$ symmetry is not broken spontaneously;
the physical chiral spectrum remains massless;
all the unwanted mirror fermions have become massive;
and  no other undesired massless states have emerged in the process.
The continuum limit of the reduced model in the SMG phase should be
a theory of free massless chiral fermions, because,
if in a {\em continuum} chiral gauge theory we turn off the gauge field,
we are left with a set of free massless chiral fermions.
Turning the gauge field back on will
then result in a lattice regularization of the desired chiral gauge theory.

The reduced model has a phase diagram spanned by the coupling constants
$g_1,g_2,\ldots$, of the interactions that we introduce
with the aim of decoupling the doublers.  In the free theory limit
$(g_1,g_2,\ldots)=(0,0,\ldots)$, the reduced model
must have a spectrum of Dirac fermions.
Why?  The hamiltonian (or lagrangian) of a free lattice fermion has a conserved
fermion number symmetry, and associated with it a conserved current.\footnote{%
  The exception is Majorana fermions.  However, these can belong only
  to real representations of any compact Lie group $G$,
  hence they do not play any role in chiral gauge theories.
}
The divergence of this current is zero on the lattice, and will remain so
in the continuum limit; the current cannot develop an anomalous divergence
in any correlation function.
The straightforward way for the lattice regularization to guarantee
the absence of an anomalous divergence in the continuum limit
is to assemble all the fermion states into Dirac fermions with respect
to the fermion number symmetry.  The same argument applies to
the conserved current of any other continuous global symmetry of the theory.
These observations were first made by Karsten and Smit,
who also derived the simplest no-go theorem in one spatial dimension \cite{KS}.

A little later, a much more powerful no-go theorem was proved by
Nielsen and Ninomiya \cite{NN}.  One considers a free lattice hamiltonian
with a compact global symmetry.  Under this symmetry,
every fermion field is endowed with a set of discrete-valued quantum numbers,
and the hamiltonian can be diagonalized in each charge sector separately.
The other requirements of the Nielsen-Ninomiya (NN) theorem are the following:
lattice translation invariance,
which implies that the momentum takes values in a periodic
Brillouin zone; a relativistic low-energy spectrum, which implies
that every massless fermion is unambiguously either right-handed (RH)
or left-handed (LH);\footnote{%
See Sec.~\ref{nogo} for the definition of the handedness.}
and finally, as a function of the momentum, the hamiltonian
must have a continuous first derivative.  Under these assumptions,
the NN theorem asserts that there is an equal number of RH and LH
massless fermions in every charge sector.\footnote{%
  For no-go theorems in euclidean space see Refs.~\cite{LK,Peli2}.
  For a no-go theorem that applies for a more general notion
  of conserved charges see Ref.~\cite{Friedan}.
}

The SMG paradigm envisages that strong-enough interactions have been turned on
to achieve an SMG phase with the properties described above.
At face value, the presence of these interactions ensures
that the NN theorem---which is a theorem about free hamiltonians---is safely
inapplicable.  Hence, the massless spectrum in an SMG phase might,
in principle, escape the conclusion of the NN theorem, and be chiral.

However, a closer look reveals another facet of the reduced model which comes
very close to satisfying the assumptions of the NN theorem.
The continuum limit of the reduced model is required to be a theory
of free massless fermions, reproducing the fermion spectrum of
the target chiral gauge theory, so that the latter will be recovered
when the gauge interactions are turned back on.  But this means that
close to the continuum limit of the reduced model,
its fermionic massless states must be almost free.
They can only interact weakly, via irrelevant interactions,
that will automatically die out in the continuum limit.
This state of affairs takes us almost all the way back to the arena
of free bilinear hamiltonians, to which the NN theorem applies.
It is thus quite natural that some generalization of the NN theorem
would be applicable in the interacting reduced model as well.
Whether or not such a generalized no-go theorem is powerful enough
to exclude a chiral fermion spectrum at a particular point
in the phase diagram of the reduced model
will then depend on the precise conditions of the theorem.

Such a generalization of the NN theorem was derived by one of us
in Refs.~\cite{NNYSPRL,NNYSlong}.  The key step is to identify
a one-particle lattice hamiltonian $\Heff(\myvec{p})$ as the inverse
of a suitable hermitian fermion two-point function $\car(\myvec{p})$
at zero frequency,
\begin{equation}
\label{Heff}
\Heff(\myvec{p}) = \car^{-1}(\myvec{p}) \ ,
\end{equation}
defined for {\em all} momenta $\myvec{p}$ in the Brillouin zone.
The precise definition of $\car(\myvec{p})$ will be given later.
If the hamiltonian (or euclidean action) of the underlying reduced model
is local, one expects that $\car(\myvec{p})$ will be
an analytic function of the momentum everywhere in the Brillouin zone,
except at {\em degeneracy points}.
A degeneracy point $\myvec{p}_c$ is a point in the Brillouin zone
where $\car(\myvec{p})$ receives a contribution from intermediate states
with $E(\myvec{p})\to 0$ for $\myvec{p}\to \myvec{p}_c$.
In this paper, we will prove the analyticity
of $\car(\myvec{p})$ away from the degeneracy points under certain assumptions
about the field content and interactions of the reduced model.\footnote{%
  The term ``degeneracy point'' reflects the fact that any state
  with vanishing energy is degenerate with the second-quantized vacuum.
}
We further distinguish between two types of degeneracy points,
which we will refer to as {\em primary singularities}
and {\em secondary singularities}.
A primary singularity $p_c$ is defined as a point where
$\car(\myvec{p}) \to \infty$ for $\myvec{p}\to \myvec{p}_c$,
which implies that $\Heff(\myvec{p}_c)$ has a zero eigenvalue.\footnote{%
  We define $\Heff(\myvec{p})$ at the primary singularities
  by demanding continuity (see Sec.~\ref{nogo}).
}
Any other degeneracy point is a secondary singularity.
By definition, at a secondary singularity
$\car(\myvec{p})$ is finite for $\myvec{p}\to \myvec{p}_c$,
hence $\Heff(\myvec{p})$ does not have any zero eigenvalues
in the vicinity of $\myvec{p}_c$.
A primary singularity requires the presence of a massless
single-particle intermediate state, whereas for a secondary singularity
any single-particle intermediate state must be gapped.
The non-analyticity at a secondary singularity arises from
intermediate states containing three or more massless fermions.

How does $\Heff(\myvec{p})$ fare with the assumptions
of the NN theorem? First, obviously, $\Heff(\myvec{p})$ is defined over
the same Brillouin zone as the reduced model.  As in the free case,
it is also subject to similar symmetry constraints, since we assume that
we are in a phase where the (to be gauged) exact global symmetry $G$
is not broken spontaneously.\footnote{%
  For the situation in one spatial dimension see Sec.~\ref{nogo}.}
Next, as explained above,
we require that the massless fermionic asymptotic states are relativistic,
and subject to irrelevant interactions only,
and that there are no massless bosons.  These assumptions will allow us
to establish that $\Heff(\myvec{p})$ has a continuous first derivative
at all degeneracy points, both primary (see Sec.~\ref{nogo})
and secondary (see App.~\ref{secondary}).
This, in turn, implies a one-to-one correspondence between the primary
singularities and the massless fermion asymptotic states.\footnote{%
  Physically, a continuous first derivative at a primary singularity
  means that the velocity of the particle is well-defined
  for $\myvec{p}\to \myvec{p}_c$.}
Everywhere else in the Brillouin zone,
$\Heff(\myvec{p})$ will be an analytic function of the momentum,
just like $\car(\myvec{p})$, with one important exception.

Let us introduce the notion of a {\em zero} of $\car(\myvec{p})$.
A point $\myvec{p}_0$ in the Brillouin zone is a zero of $\car(\myvec{p})$
if the latter has a zero eigenvalue for $p=p_0$.
By Eq.~(\ref{Heff}), a zero of $\car(\myvec{p})$
turns into a pole of $\Heff(\myvec{p})$.  Thus, if $\car(\myvec{p})$ has
a nonempty set of zeros, then $\Heff(\myvec{p})$ will not have
a continuous first derivative at the same points.  As a result,
a crucial condition of the NN theorem will be violated,
and the theorem will not apply.

The applicability of the NN theorem thus relies on the ability
to construct fermion two-point functions that do not have any zeros.
This brings us to the following issue:
the identification of a suitable set of lattice operators that will serve
as interpolating fields for the fermionic asymptotic states
of the reduced model at a given point of its phase diagram.
It is convenient to define a {\em complete set} of interpolating fields
in a given charge sector by the following two requirements:
(1) The massless fermion asymptotic states
are in one-to-one correspondence with the primary singularities
of $\car(\myvec{p})$ for this set of interpolating fields;
(2) $\car(\myvec{p})$ is free of zeros.

In practice, building a complete set of interpolating fields can be
a trial and error process, which could be helped by hints about the dynamics.
First, it is quite natural for a lattice fermion field to interpolate
more than one massless state, or, equivalently, to have several primary
singularities.  The simplest example is a naive fermion
on a one-dimensional spatial lattice.  This is a single-component field
with the dispersion relation $E=\frac{1}{a}\sin(ap)$,
where $a$ is the lattice spacing.  The massless spectrum\footnote{%
  In one spatial dimension, with no concept of helicity,
  massless fermions can be right-moving or
  left-moving.   The abbreviations RH and LH when used in the context
  of models in one spatial dimension will be taken to refer to right-movers
  and left-movers, respectively, throughout this paper.
}
then consists of a RH state at $p_c=0$ and a LH state at $p_c=\p/a$.
As shown in App.~B of Ref.~\cite{NNYSlong}, attempting to build a
one-to-one correspondence between these massless fermion asymptotic states
and the interpolating fields results in an ``over-complete'' set
of interpolating fields that suffers from the presence of zeros.

Notice that our definition of a complete set of interpolating fields
does not refer explicitly to any gapped fermion asymptotic states.
The reason is that gapped states (whether relativistic or not)
do not generate singularities at zero frequency.
The primary singularities of $\car(\myvec{p})$,
and thus the zeros of $\Heff(\myvec{p})$,
correspond to massless fermions only. As a result,
both the original NN theorem and its generalization discussed here
are ``oblivious'' to the presence of any gapped fermion states.

This state of affairs is convenient, in that it means that it does not matter
if our set of interpolating fields generates any gapped fermion states
(in addition to the massless fermion states),
so long as it satisfies the requirements of a complete set defined above.
However, the requirement that $\car(\myvec{p})$ should have no zeros
indirectly constrains how Dirac fermions can be interpolated.
A massive Dirac fermion consists of a RH and a LH component which are coupled
via  the mass term.  As we will explain in detail below, if both chiralities
are interpolated, this should not lead to propagator zeros.  However,
if only one of the two chiralities is interpolated, while the other is not,
this leads to the unwanted appearance of a {\it kinematical zero}
in $\car(\myvec{p})$.
These kinematical zeros can be avoided by simply adding
to the set of interpolating fields new fields that will interpolate
the missing chirality component of the relevant Dirac fermions.

A case where a propagator zero is unavoidable is when our starting point
is a bilinear hamiltonian or action that has a built-in pole.
Such a (euclidean) action was proposed
long ago by Rebbi \cite{Rebbi} in an early effort to put chiral gauge theories
on the lattice.  However, it was soon realized that a pole in the action
acts as a ghost state \cite{Peli2,Peli1}.  In particular, in four dimensions
it contributes to the one-loop beta function with the same magnitude
as a fermion field, but with an opposite sign.  We have recently generalized
this conclusion using an effective low-energy framework \cite{zeros}.

As the example of Rebbi's proposal shows, generating a pole in the action
requires a highly non-local kinetic term.  If, on the other hand,
the underlying theory is local, it is unlikely that the severe nonlocality
needed for generating such ghost states would emerge at the level
of the low-energy effective theory \cite{Youetal}.  Turned around,
this observation suggests that any propagator zeros encountered
in the SMG phase of a local reduced model should be kinematical zeros,
which are removable by a judicious choice of the interpolating fields.

Our current understanding of the issue of propagator zeros,
which has evolved considerably since our earlier paper \cite{zeros},
is the following.  We conjecture that the same SMG interactions
that gap the mirrors also generate opposite-chirality bound states.
These bound states pair up with the elementary mirror fermions,
thereby forming massive Dirac fermions.  Hence, in order to obtain
a complete set of interpolating field free of propagator zeros,
one should add to the set of elementary fields, which
interpolate the mirror fermions, a suitable set of composite fields that will
interpolate the opposite-chirality bound states.

A popular testbed of the SMG paradigm is the so-called 3450 model
\cite{CGP,Yoshio,WWPRB,DMW,ZZWY}.
The target chiral gauge theory is an abelian theory in one spatial dimension,
with two (say) LH fields with charges 3 and 4, and one RH field with charge 5.
Since $3^2+4^2=5^2$, this fermion content satisfies
the anomaly cancellation condition in two spacetime dimensions.
When the SMG paradigm is applied to the 3450 model, the starting point
is an abelian lattice theory with massless Dirac fermions with charges
3, 4, and 5, representing a doubled spectrum.
One then aims to find an SMG phase where the mirrors have become
massive, while the fermion spectrum of the target chiral gauge theory
remains massless.  The reason for the name ``3450 model'' is that a neutral
``spectator''
Dirac fermion is added to the lattice theory.  Having a zero charge,
this fermion field does not interact with the gauge field.
But it is employed in the construction of the interaction terms that, one hopes,
would generate the SMG phase in which all the doublers become massive,
and one is left with a massless fermion spectrum consisting of LH states
with charges 3 and 4, and RH states with charges 5 and 0.

An attempt to decouple the mirrors was made in Ref.~\cite{CGP}, where
a concrete reduced-model realization of the 3450 model was proposed.
The goal of Ref.~\cite{CGP} was to obtain an SMG phase by introducing
additional scalar fields that interact strongly with the fermion fields
via Yukawa couplings.
As a probe of the fermion spectrum, the vacuum polarization diagram
was calculated in the reduced model, \ie, the two-point function
of the conserved current of the global U(1) symmetry to be gauged
in the full model.
In two space-time dimensions, a massless fermion intermediate state generates
a directional discontinuity at zero momentum.  In units of the discontinuity
created by a single chiral fermion of unit charge, the strength of the total
discontinuity is $\sum_i q_i^2$, where the sum runs over the U(1) charges $q_i$
of all the massless chiral fermions, both RH and LH.
An undoubled massless spectrum would thus exhibit a discontinuity of
total relative strength $3^2+4^2+5^2=50$, coming from the massless
LH fermions with charges 3 and 4, and the RH fermion with charges 5.

In Ref.~\cite{CGP} the discontinuity of the vacuum polarization diagram
was calculated numerically in the mirror sector
at a single point inside the would-be SMG phase.
Instead of the desired result, which is a vanishing discontinuity in the
mirror sector, the actual result was consistent with a discontinuity
of total relative strength of 50.  Clearly,
the simplest interpretation of this result is that the mirror fermions
remained massless, and thus that the spectrum remained doubled,
and not chiral.\footnote{%
  See Ref.~\cite{CGP} for the systematic uncertainties involved
  in the calculation.  See also Ref.~\cite{Yoshio} for an attempt
  to explain the failure of Ref.~\cite{CGP}.
}
As pointed out in Ref.~\cite{CGP}, an alternative interpretation would be
the appearance of a new massless Dirac fermion of charge 5,
whose LH component is the original mirror state of the fermion field
with charge 5, while its RH component is a bound state
of the fermion and scalar fields of the model.
Such a situation would imply that the total fermion spectrum
in each charge sector remains chiral.
We comment that while this situation cannot be ruled out
based on the result of Ref.~\cite{CGP} alone, it is unlikely in our opinion,
even if we set aside the generalized no-go theorem which is the main topic
of the present paper.  The reason is that the model of Ref.~\cite{CGP} contains
no symmetry that would protect the masslessness of such a Dirac fermion,
and so it would be a remarkably fine-tuned situation
if its mass happened to be zero
(within the uncertainties of the calculation) right at the point
in the phase diagram where the numerical calculations were carried out.

Recently, building on heuristic arguments presented in Refs.~\cite{WWPRB,DMW},
a different reduced-model realization of the 3450 model was put forward
in Ref.~\cite{ZZWY}, which we will refer to as the \ZZWY\ model.
Introducing two judicially chosen 6-fermion interactions,
the successful development of an SMG phase with gapped mirror
fermions was announced.  Since this result is in apparent conflict with
the generalization of the NN theorem to interacting
(reduced) models, our goal in this paper is to revisit the considerations
of the generalized theorem \cite{NNYSPRL,NNYSlong},
using the \ZZWY\ model as a laboratory whenever possible.

This paper is organized as follows.  We begin in Sec.~\ref{bounds}
with a fresh examination of the issue of propagator zeros, revisiting
our discussion in Ref.~\cite{zeros}.  We first explain why propagator zeros
tend to arise when mirror fermions are gapped.
Because the underlying reduced model is local by assumption,
we now realize that the propagator zeros are unlikely
to represent ghost states \cite{Youetal}, and are thus more likely
to be kinematical in nature.  We then
introduce the (conjectured) bound-state formation mechanism:
The same SMG interactions that gap the mirrors also generate opposite-chirality
bound states, which, in turn, pair up with the elementary mirror fermions
to form massive Dirac fermions free of propagator zeros.\footnote{
  While in Ref.~\cite{zeros} we noted that the presence of bound states
  could in principle affect our conclusions, we did not appreciate
  that bound-state formation would turn out to play a key role.
}
We give explicit formulae for the composite operators that would create
the bound states in the \ZZWY\ model, and should thus
be added to the set of interpolating fields.

In Sec.~\ref{nogo} we turn to the generalized no-go theorem.
In contrast with Refs.~\cite{NNYSPRL,NNYSlong} where analyticity properties
were essentially postulated, in this paper we prove the analyticity of
$\car(\myvec{p})$ for reduced models defined by a local hamiltonian
that depends on fermion fields only (as is the case for the \ZZWY\ model).
We discuss the remaining conditions of the generalized theorem,
which include the absence of zeros in $\car(\myvec{p})$,
and the assumption that the continuum limit of the reduced model
is a theory of relativistic free massless fermions, with no additional massless
bosonic states.  Technical details are mostly relegated to App.~\ref{Rp}.
Finally, we briefly explain how the gauge-fixing approach
to the construction of four-dimensional lattice chiral gauge theories
evades the generalized no-go theorem.

In Sec.~\ref{slimit} we prove an independent, but related, simple theorem.
We consider a reduced model in which a subset of the
fermion degrees of freedom participates in some strong interactions.
We then show that when the (uniform) strong-coupling limit is taken,
the fermion fields split into two decoupled sectors.  In particular,
the fermion degrees of freedom that do not participate in
the strong interactions then form a decoupled free (or weakly coupled) theory,
which is thus subject to the NN theorem.  We list several examples,
including in particular the application of this theorem to the \ZZWY\ model.

In Sec.~\ref{nogo2dim} we turn to the implications of the
generalized no-go theorem for two-dimensional theories.
Two-dimensional theories are, in a sense, more complicated than
four-dimensional theories, because many more relevant or marginal
operators exist in two dimensions, and their role needs to be
understood.  We confront the generalized no-go theorem with the properties
of two-dimensional SMG reduced models in general,
and with the known features of the ZZWY model in particular.
While the SMG interactions themselves are always chosen to be irrelevant,
they can induce four-fermion interactions without derivatives that
respect all the symmetries of the SMG reduced model.
In two dimensions, such induced four-fermion interactions are renormalizable,
hence they can have a profound impact on the resulting continuum theory.

Our work can only have tentative implications for any specific SMG model;
for the generalized no-go theorem to apply, all of its assumptions
need to be satisfied, and this must be checked on a case by case basis.
In Sec.~\ref{roadmap} we use our analysis to compile a list of open questions.
We focus on the fate of propagator zeros and on the applicability of
each assumption of the generalized no-go theorem.
These key questions will have to be sorted out in the ZZWY model, as well as
in any other SMG model, in order to determine whether or not it succeeds
in recovering a chiral gauge theory in the continuum limit.

We end with some concluding remarks in Sec.~\ref{conc}.
Appendix~\ref{HZZWY} provides details on the \ZZWY\ hamiltonian,
focusing mainly on its bilinear part, including
the impact of the strong-coupling limit of Sec.~\ref{slimit},
as well as a few details on the interaction hamiltonian and the symmetries.
Proofs of most of the properties of $\car(\myvec{p})$
needed for the generalized no-go theorem are relegated to App.~\ref{Rp}.

\section{\label{bounds} SMG, propagator zeros, and bound-state formation}
The goal of the SMG paradigm is to gap the mirror chiral component
of a Dirac fermion while leaving massless the other chirality,
as it is part of the chiral spectrum of the target chiral gauge theory.
In order to see why the gapping of the mirror component tends to lead
to the appearance of a propagator zero, consider the example
of a continuum massless Dirac fermion, with propagator\footnote{%
  Our conventions are as follows.  We use $d$ for the number of
  spatial dimensions.  We denote $d$-vectors as, \eg, $\myvec{x}$, $\myvec{p}$,
  and their inner product as $\myvec{x}\cdot\myvec{p}$.
  For quantities that carry spacetime indices
  we use Minkowski space conventions with $\eta_{\m\n}=diag(1,-1,-1,-1)$.
  For example, $p^2 = p^\m p_\m = \o^2 - \myvec{p}^2$ where $\o=p_0$.
}
\begin{equation}
\label{m0pole}
\frac{\sl{p}}{p^2} \ .
\end{equation}
Chiral symmetry follows from the fact that this propagator anti-commutes
with $\g_5$.
If, for example, the RH component has been gapped somehow,
while the LH component remains massless, then the RH pole must have moved
away from zero, so that the same two-point function would now take
the new form
\begin{equation}
\label{pzero}
P_L \frac{\sl{p}}{p^2} P_R + P_R \frac{\sl{p}}{p^2-m^2} P_L \ .
\end{equation}
where $P_{L,R}$ are the chiral projectors.
While the last term indicates
the presence of a massive state, this propagator now has a RH zero instead
of the original massless pole.\footnote{
  A propagator zero occurs if mixed-chirality regular terms are
  added to Eq.~(\ref{pzero}) as well.
}

Assuming that propagator zeros develop in an SMG phase, what is
their physical significance?  The dynamics of an SMG phase might not be
easily tractable because SMG requires strong interactions.
Before we turn to this question, let us consider another example
taken from a free continuum theory.
This time we start from the massive Dirac propagator
\begin{equation}
\label{Dprop}
G = \frac{\sl{p}-m}{p^2-m^2} \ ,
\end{equation}
which has a pole at $p^2=m^2$, and no zeros.  Let us now pick out
the RH chirality of this Dirac fermion by applying suitable
projectors on the two sides of $G$. The result,
$P_R G P_L$, readily coincides with the rightmost term in Eq.~(\ref{pzero}).
This shows that a propagator zero can be a ``mundane'' kinematical singularity
resulting from the application of a projection to an ordinary massive
Dirac propagator.

As we have discussed in the introduction, the alternative is
an ``irremovable'' propagator zero that arises from a nonlocal action
with a pole in its bilinear part, which in turn represents a ghost state.
However, the idea is that the SMG phase develops in a theory
where, as a rule, the underlying lagrangian (or hamiltonian) is local
by construction.  Hence, it is unlikely that the kind of nonlocality
needed for ghost states would develop at the level of the
effective theory that controls the long-distance behavior \cite{Youetal}.

The propagator zeros found in an SMG phase are thus more likely to be
kinematical singularities.  This means that it must be possible
to reproduce each propagator zero by applying chiral projectors to a pertinent,
massive Dirac propagator, as in the simple example we have just considered.
But this conclusion immediately raises another question.
If the propagator zero can be reproduced by projecting out the RH component
of a massive Dirac fermion found in the spectrum of the theory,
then, evidently, the corresponding Dirac field
must have a LH component as well.  However, the LH component of this massive
Dirac field must be different from the LH component of the original
massless Dirac fermion.  The reason is that, first, the SMG interactions
are constructed to involve only the RH component of the original
massless Dirac fermion (any residual coupling to the LH component
is suppressed by design).  Second, if the goal of the SMG program
is to be achieved, then the LH component of the original massless fermion
has to remain massless.  The LH component of the massive Dirac fermion
should thus be supplied by the SMG dynamics itself,
in other words, it must arise as a bound state in the SMG phase!
A constraint is that this bound state transforms in the
same representation of $G$ as the gapped RH component of the
original Dirac fermion, in order to avoid spontaneous breaking of
the symmetry group $G$.

We will now argue that, for each Weyl fermion that has been gapped
in an SMG phase, the SMG dynamics can in principle generate an
opposite-chirality bound state.  Ultimately, the massive spectrum in the
SMG phase would thus consist of ``hybrid'' Dirac fermions,
each of which has one chirality component
which is elementary, while the other chirality component is a bound state.
The full propagator of each massive Dirac fermion is free of zeros.

\begin{figure}[t]
\begin{center}
\includegraphics*[width=10cm]{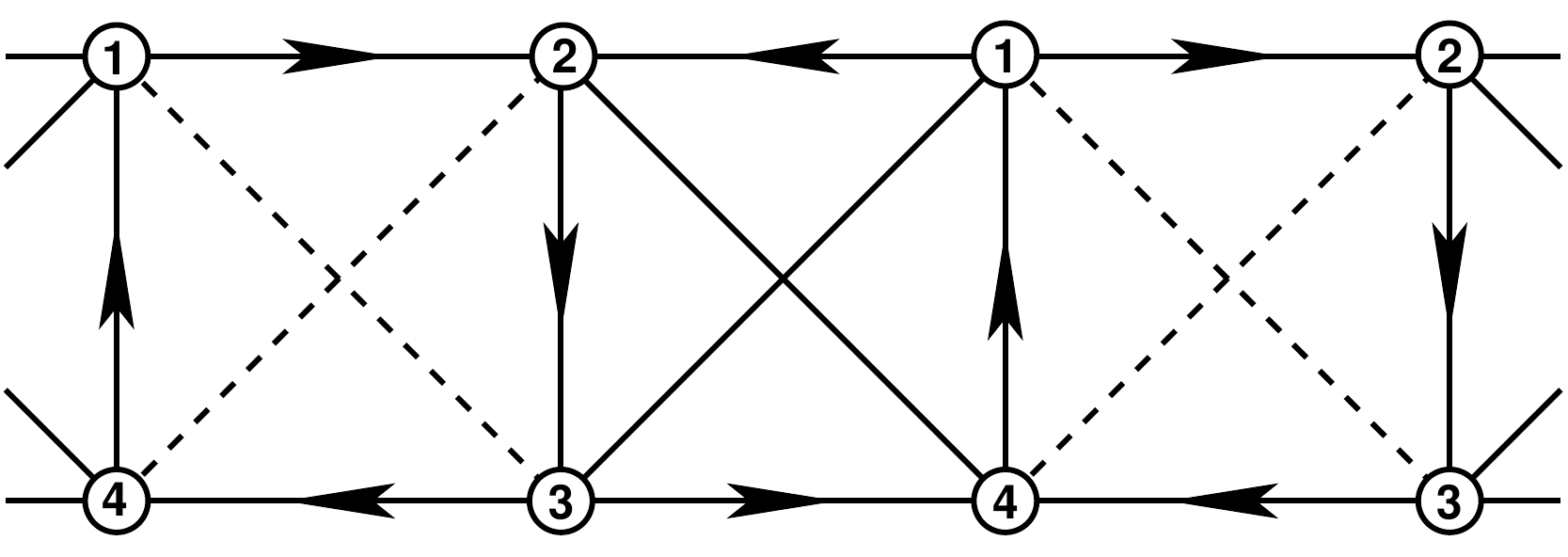}
\end{center}
\begin{quotation}
\floatcaption{3450latt}%
{The lattice of the 3450 model of Ref.~\cite{ZZWY}.  The lattice shown
in Fig.~1(a) of that paper is here rotated by $90^0$ clockwise,
so that the physical space direction is horizontal.
The lattice consists of two inter-connected one-dimensional chains.
Edge~A (on the left in Fig.~1(c) of Ref.~\cite{ZZWY}) is here the upper chain,
while edge~B is the lower chain.  The directional links have complex coupling
$\t_1 = t_1 e^{i\frac{\p}{4}}$, while the undirectional links have real coupling
$t_2$ (solid line) or $-t_2$ (dashed line).  The unit cell of this lattice
is a $2\times 1$ rectangle (in lattice units), and the numbers inside
the circles represent different sublattices.  To avoid confusion,
note that the sublattice index is different from the species
index $I$ in Eq.~(\ref{H0}).  For more details, see App.~\ref{HZZWY}.
}
\end{quotation}
\vspace*{-4ex}
\end{figure}

As we will see, the mechanism of bound-state formation is very general.
Nevertheless, to make the discussion more concrete we will consider here
the ZZWY lattice construction for the 3450 model \cite{ZZWY}.
We thus pause to describe the main features of this model.
More details may be found in App.~\ref{HZZWY}.

The \ZZWY\ model is defined by a lattice hamiltonian.
The time coordinate remains continuous while space is discretized.
The hamiltonian consists of a free part $H_0$ and an interacting part $\Hint$.
There are four single-component fermion fields $\j_I$, where the
``species'' index $I=1,2,3,4$ corresponds to charges 3, 4, 5 and 0
under the U(1) symmetry to be gauged.\footnote{
  The gauge field is turned off in the \ZZWY\ model, \ie,
  the reduced model is considered in Ref.~\cite{ZZWY}.
}
The fermion fields live on the lattice shown in \Fig{3450latt},
which consists of two interconnected one-dimensional chains.
The free hamiltonian is given by
\begin{equation}
\label{H0}
H_0 = \bj_1 \ch(\t_1,t_2) \j_1 + \bj_2 \ch(\t_1,t_2) \j_2
      + \bj_3 \ch(\t_1^*,t_2) \j_3 + \bj_4 \ch(\t_1^*,t_2) \j_4 \ ,
\end{equation}
where $\t_1$ is a complex parameter while $t_2$ is real
(for the precise form of $\ch(\t_1,t_2)$ and the actual values
of $\t_1$ and $t_2$, see App.~\ref{HZZWY}).  $\ch(\t_1,t_2)$ supports
one LH chiral mode on edge~A, and one RH chiral mode on edge~B.
For $\ch(\t_1^*,t_2)$ the chiralities are reversed.
The spectrum of the target 3450 chiral gauge theory consists of LH fermions
with charges 3 and 4, and RH fermions with charges 5 and 0.
The physical chiral mode of each lattice species thus always lives on edge~A,
while the opposite chirality mode on edge~B is always the doubler,
or mirror fermion.  The goal of the SMG program is to gap
all the edge-B chiral modes, without breaking spontaneously the U(1) symmetry.
Below, it will often be convenient to distinguish between the fermion fields
on the two edges using a separate notation: $\x_I$ for the edge-A fields,
and $\c_I$ for the edge-B fields.

We next consider the interaction hamiltonian, which takes the form
\begin{equation}
\label{Hint}
\Hint = g_1 H_1 + g_2 H_2 \ .
\end{equation}
Here $g_1,g_2$ are coupling constants, while $H_1,H_2$ are (lattice sums of)
local 6-fermion operators.  The interaction hamiltonian couples the edge-B fermions
only, $\Hint=\Hint(\c_I)$.  Notice that the bilinear hamiltonian $H_0$
has a separate U(1) fermion number symmetry for every fermion species.
Two linear combinations of these U(1) transformations are broken explicitly
by $\Hint$, while two other linear combinations remain as exact global
symmetries of the full hamiltonian.  One of them, of course, corresponds
to the gauge symmetry of the 3450 model.  For more details on
the symmetries, as well as on the interaction terms, see App.~\ref{syms}
or Ref.~\cite{ZZWY}.

Introducing interpolating fields for the (anticipated) bound states,
\begin{equation}
\label{BaI}
\cb_{I,a}(x) = \frac{1}{6}\frac{\d H_a}{\d \c_I^\dagger(x)} \ ,
\qquad a=1,2\ ,
\end{equation}
where the coordinate $x$ labels the edge-B sites, and their sum
\begin{equation}
\label{BI}
\cb_I(x) = g_1 \cb_{I,1}(x) + g_2 \cb_{I,2}(x) \ ,
\end{equation}
we can reexpress the interaction hamiltonian as
\begin{equation}
\label{HBI}
\Hint = \sum_x \sum_{I=1}^4 \Big(\c_I^\dagger(x) \cb_I(x) + \hc \Big) \ .
\end{equation}
Next, introducing two-component fermion fields
\begin{eqnarray}
\label{bsD}
\dirac_I = \left(\begin{array}{c} \c_I \\ \cb_I \end{array} \right) \ ,
\\
\bdirac_I = \rule{0ex}{3ex}
\left(\begin{array}{cc} \cb^\dagger_I & \c^\dagger_I \end{array} \right) \ ,
\end{eqnarray}
it follows that the interaction hamiltonian takes the form of a mass term
for the (two-dimensional) Dirac fermions $\dirac_I(x)$, explicitly,
\begin{equation}
\label{HintM}
\Hint = \sum_x \sum_{I=1}^4 \bdirac_I(x) \dirac_I(x) \ .
\end{equation}
The hamiltonian does not contain explicit kinetic terms
for the bound-state fields $\cb_I(x)$, but these may be generated dynamically.

Finally let us return to the issue of propagator zeros.
Let us assume that near some critical momentum $p_0$, the propagator
of the $\dirac_I$ field behaves like a massive Dirac propagator,
\begin{equation}
\label{propbs}
\svev{\dirac_I\,\bdirac_I} = \frac{\o\g_0 - (p-p_0)\g_1 - m_I}
     {\o^2 - (p-p_0)^2 - m_I^2} + \cdots \ ,
\end{equation}
where the ellipsis stand for terms suppressed by the lattice spacing,
which we will disregard in this section.
Here $\g_0,\g_1$ satisfy the two-dimensional Dirac algebra and, moreover,
we assume that the chirality matrix $\g_0\g_1$ is equal to the
third Pauli matrix $\s_3$ in the basis where $\dirac_I$ is given by Eq.~(\ref{bsD}).
The two-dimensional chiral projectors are thus $P_{R,L} = \half(1\pm \s_3)$.
If we now consider only the $\c_I$ component, we obtain the projected propagator
\begin{equation}
\label{propchi}
\svev{\c_I\,\c_I^\dagger} = P_R \svev{\dirac_I\,\bdirac_I} P_L
= P_R\, \frac{\o\g_0 - (p-p_0)\g_1}{\o^2 - (p-p_0)^2 - m_I^2} \,P_L \ .
\end{equation}
Focusing on $\o=0$, which will be the relevant case for the discussion
in the next section, this projected propagator has a zero eigenvalue
for $p-p_0=0$.
This demonstrates how the mechanism
of bound-state formation can turn the propagator zero of Eq.~(\ref{propchi}),
anticipated in the SMG phase, into a component of the propagator of a normal,
massive Dirac field, Eq.~(\ref{propbs}).  As a result, the ghost states
associated with an ``irremovable'' propagator zero are avoided,
and the effective theory associated with the long-distance
degrees of freedom can be local.

We comment that one can allow for independent
renormalization factors, $Z_t$ and $Z_s$,
for the temporal and spatial components in Eq.~(\ref{propbs}).
Any deviation of the ratio $Z_s/Z_t$ from unity can be absorbed
into a rescaling of the lattice spacing, while a common $Z_s=Z_t$
represents as usual an overall wave-function renormalization factor
for the fermion field occurring in the two-point function.

Starting from the explicit form of the interaction hamiltonian
of the \ZZWY\ model, we have
identified candidate composite fields which can serve as
interpolating fields for the opposite-chirality bound states, that, in turn,
should pair up with the gapped elementary chiral fermions in order to form
massive Dirac fermions in the SMG phase.  Clearly, this construction
is very general, and could be applied to a wide range of SMG models.
In contrast, proving that the said bound states actually exist
is a more difficult task that must be addressed on a case-by-case basis.
As a rule, we expect that the existence of the bound states
can be proved when a strong-coupling expansion is available.
Relevant examples include the strong-coupling phase
of the Eichten-Preskill model \cite{EP}, analyzed in Ref.~\cite{GPR}, as well as
the qualitatively similar example presented in Ref.~\cite{Youetal}.
Additional examples of propagator zeros were discussed in Refs.~\cite{YWOX,SYX}.
In particular, Ref.~\cite{YWOX} employed a strong-coupling expansion
similar to that of Refs.~\cite{Youetal,GPR}.  The formation of bound states
within the strong-coupling expansion was proved in Ref.~\cite{GPR},
but was not explicitly considered in Refs.~\cite{Youetal,YWOX,SYX}.

As for the \ZZWY\ model itself, no such strong-coupling expansion
is available.  The technical reason why such an expansion is not feasible
is that the multi-fermion interactions contain hopping terms.
In order to study whether or not the conjectured bound states exist
one must therefore resort to numerical methods.

Let us recap the (conjectural) situation with regard to zeros of $\car(p)$
in the ZZWY model.
The mechanism of bound-state formation offers a natural escape route
from the ghost states that would otherwise be associated with propagator zeros
in the SMG phase.  We conjecture
that the set $\{\x_I,\c_I,\cb_I\}$ forms a complete set of interpolating fields.
Apart from the original elementary fermion fields of the model:
the edge-A fields $\x_I$ and the edge-B fields $\c_I$,
this set includes composite fields $\cb_I$.
The fields $\{\c_I,\cb_I\}$ can pair up to form
massive Dirac fermions in the SMG phase.  The addition of the
composite fields $\cb_I$ thus eliminates the propagator zeros
found if we use only the edge-A and edge-B elementary fields.

Once the two-point functions of the set $\{\x_I,\c_I,\cb_I\}$ are free
of propagator zeros, and provided that the additional conditions stated
in the next section are satisfied, the generalized no-go theorem applies.
Then the spectrum in each charge channel must be vector-like,
and thus new doublers must appear.
Notice that the theorem does not provide any information about
which interpolating fields from our complete set generate massless states,
nor on the location of any new degeneracy points that might occur
in the Brillouin zone.
One possibility that is clearly compatible with the mechanism
of bound-state formation is that the elementary edge-B fields $\c_I$
together with the composite fields $\cb_I$ generate massive states only, and
the new doublers  are generated by the edge-A fields $\x_I$.
In this case it may turn out that the smaller set that consists of
the edge-A fields $\x_I$ only already comprises a complete set
of interpolating fieds free of propagator zeros in the SMG phase.
Such a behavior would in fact resemble the spectrum of the edge-A fields
in the strong-coupling limit discussed in Sec.~\ref{slimit}
and App.~\ref{HpSlim} below.

We stress, however, that we have no concrete knowledge about
the actual situation, and that other scenarios might be possible as well.
One such alternative scenario is that the new massless doublers of the
edge-A massless fermions arise in the SMG phase as bound states interpolated
by yet another set of composite fields, denoted $\ca_I$. In this case,
the complete set of interpolating fields would take the form
of $\{\x_I,\c_I,\cb_I,\ca_I\}$.  The rest of the argument is unchanged:
once we have succeeded in constructing a complete set of interpolating fields
with two-point functions free of propagator zeros,
the no-go theorem can be applied.

What if it turns out to be impossible to build any complete set of
interpolating fields, whose two-point functions are free of propagator zeros?
While we cannot rule out this option,
we have given in Ref.~\cite{zeros} strong arguments that such irremovable zeros
are ghost states, that render the SMG phase of the theory inconsistent.
But as we have already stressed in this paper, it is unlikely that the kind
of nonlocality needed for ghost states would develop in the SMG phase
of any reduced model with a local action \cite{Youetal}.
We thus expect that it should always be possible to find a complete set
of interpolating fields whose associated $\Heff$ is free of poles.

\section{\label{nogo} Applicability of the Nielsen-Ninomiya theorem}
The previous section suggests the following conjectural physical picture.
In an SMG phase of a local reduced model, propagator zeros are likely to be
kinematical singularities.  They arise because the gapped mirror components
of the elementary fermion fields combine with
opposite-chirality bound states to form massive Dirac fermions.
As a result, using the elementary fermion fields only as the set of
interpolating fields effectively projects onto a single chirality of these
massive Dirac fermions, and the projection generates the propagator zeros.
The addition of the bound-state composite fields to the set of
interpolating fields then provides us with a complete set,
with two-point functions that are free of propagator zeros while their
primary singularities are in one-to-one correspondence
with the massless fermion asymptotic states.

In this section we present a new proof of the generalized no-go theorem
\cite{NNYSPRL,NNYSlong}, making certain assumptions
about the field content and the hamiltonian of the reduced model.
With these assumptions,
the theorem is valid in both $d=1$ and $d=3$ spatial dimensions, and
the basic framework applies anywhere in the phase diagram
of a given reduced model.  However, as we discuss later on,
some important elements depend on the properties of the SMG phase
in question, and, more generally, on whether we are in two or four dimensions.

We assume that a complete set of fermion interpolating fields
has been constructed at the point in the phase diagram under consideration.
We will use the notation $\J_a$, with $a$ as a generic index,
for all the interpolating fields that belong to this complete set,
both elementary and composite.
For reasons that will become clear in the proof of analyticity,
it is advantageous to consider the {\em retarded} and {\em advanced}
two-point functions (and not the {\em time-ordered} functions
common in the path-integral framework).
The retarded anti-commutator is defined by
\begin{equation}
\label{Retc}
R_{ab}(\myvec{x},t)=i\th(t)\,\sbra{0}\{\J_a(\myvec{x},t)\,,
\J^\dagger_b(\myvec{0},0)\}\sket{0} \ .
\end{equation}
The space and space-time Fourier transforms are (from now on
we omit the indices)
\begin{equation}
\label{Retcx}
\hat{R}(\myvec{p},t)=\sum_{\myvec{x}}
e^{ - i\myvec{p}\cdot\myvec{x}}\,R(\myvec{x},t) \ ,
\end{equation}
\begin{equation}
\label{Retcxt}
\tilde{R}(\myvec{p},\o)=\int_0^\infty dt\, e^{i\o t}
\hat{R}(\myvec{p},t) \ , \qquad \Im\o>0 \ .
\end{equation}
Excluding the degeneracy points, we also define ($\e>0$)
\begin{equation}
\label{limRetc}
\car(\myvec{p})=\lim_{\e\to 0}\tilde{R}(\myvec{p},\o=i\e) \ .
\end{equation}
Finally, $\Heff(\vec{p})$ is defined by Eq.~(\ref{Heff}).

For the advanced anti-commutator $A(\myvec{x},t)$, one replaces
$\th(t)$ by $-\th(-t)$ in Eq.~(\ref{Retc}).  The corresponding Fourier transforms
are denoted $\hat{A}(\myvec{p},t)$ and $\tilde{A}(\myvec{p},\o)$,
with now $\Im\o<0$.  The definition of $\ca(\myvec{p})$ is similar
to Eq.~(\ref{limRetc}), with $\o=-i\e$ replacing $\o=+i\e$.

\mynext
The functions $\car(\myvec{p})$ and $\ca(\myvec{p})$ defined above
have the following properties:

\mynext {\bf 1.} {\em Equality of $\car(\myvec{p})=\ca(\myvec{p})$.}
Except at degeneracy points, where there are
intermediate states with vanishing energy,
when $\o$ tends to zero from the relevant half-space of the complex plane
the common boundary value of the retarded and advanced correlators,
$\tilde{R}(\myvec{p},\o)$ and $\tilde{A}(\myvec{p},\o)$,
is equally given by $\car(\myvec{p})$ or $\ca(\myvec{p})$.

This property follows immediately by introducing a complete set
of intermediate states, and noting that the momentum dependence always
takes the form of the familiar energy denominators,
\begin{equation}
\label{Edenom}
\frac{1}{\o\pm E(\myvec{p})} \ ,
\end{equation}
where $E(\myvec{p})$ is the energy of an intermediate state
with momentum $\myvec{p}$ ({\it cf.} App.~\ref{Rpdagger}).
Hence, the limit $\o\to 0$ exists, except at the degeneracy points.

\mynext {\bf 2.} {\em $\car(\myvec{p})$ is hermitian.}
The proof is given in App.~\ref{Rpdagger}.
By Eq.~(\ref{Heff}), $\Heff(\myvec{p})$ is hermitian as well.  Strictly speaking,
$\car(\myvec{p})$ is undefined at the primary singularities,
where it diverges.  We define $\Heff(\myvec{p})$
at the primary singularities by requiring continuity, hence it is hermitian
everywhere in the Brillouin zone. Continuity at a primary singularity is
a corollary of Eq.~(\ref{relzlog}) below, which, as discussed later on, follows from
the requirement that the continuum limit of the reduced model is a theory of
free massless fermions.

\mynext {\bf 3.}  {\em Analyticity.} $\car(\myvec{p})$ is an analytic function
of $\myvec{p}$ except at the degeneracy points.
The proof is given in App.~\ref{Rpwedge}.  It assumes that the reduced model
contains fermion fields only, as well as a strong form of locality
of the hamiltonian, and it makes use of the edge-of-the-wedge theorem.

\medskip

Assuming that $\car(\myvec{p})$ has no zeros,
it follows that $\Heff(\myvec{p})$ is also an analytic function
of $\myvec{p}$ except at the degeneracy points.
What remains to be discussed is the behavior of $\Heff(\myvec{p})$
near these degeneracy points.  We will be interested mainly
in the primary singularities, which is where $\Heff(\myvec{p})$
can have zero-energy eigenstates that correspond to the massless fermion
asymptotic states.  Secondary singularities were discussed in detail
in Refs.~\cite{NNYSPRL,NNYSlong}, and we will not repeat this discussion here.
Instead, we give in App.~\ref{secondary} an example of a secondary
singularity that highlights the role of such points.

We require that the massless fermion excitations are relativistic.
This means that the leading behavior near a primary singularity $p_c$ is
\begin{subequations}
\label{relz}
\begin{eqnarray}
\label{relz1}
E &=& \pm (p-p_c) + \cdots \ , \hspace{8.0ex} d=1 \ ,
\\
\label{relz3}
H_{2\times 2} &=& \pm\,\vec\s\cdot(\vec{p}-\vec{p}_c) + \cdots \ ,
\qquad d=3 \ .
\end{eqnarray}
\end{subequations}
where the ellipsis stand for terms suppressed by powers
of the lattice spacing.\footnote{%
  As in Sec.~\ref{bounds}, in Eq.~(\ref{relz}) one can allow for a general ratio
  $Z_t/Z_s$ of renormalization factors, or equivalently, a renormalization of
  the speed of light.  Compare also Eq.~(\ref{Epma}).}
The $\pm$ signs define the chirality of the massless state,
with a plus (minus) sign for a RH (LH) state.
In words, for $d=1$ the hamiltonian must have an eigenvalue that
behaves like $\sim \pm (p-p_c)$.  In $d=3$ it must be possible
to choose a basis for the hamiltonian such that all the
eigenvectors with near-zero energy
are assembled into diagonal $2\times 2$ blocks of the form
$\sim \pm\,\vec\s\cdot(\vec{p}-\vec{p}_c)$.

Analytic corrections to the
relativistic behavior~(\ref{relz}), which come in powers of $a(p-p_c)$,
will not violate the conditions of the NN theorem.
For example, as already mentioned in the introduction,
a naive fermion in $1+1$ dimensions has $E = \frac{1}{a}\sin(ap)$.
This dispersion relation is analytic in the whole Brillouin zone,
which is topologically a circle;
and indeed, in addition to the RH massless state at $p_c=0$ there is
a LH doubler at $p_c=\p/a$, consistent with the no-go theorems
\cite{KS,NN}.

The remaining condition of the NN theorem that needs to be established
is that $\Heff(p)$ has a continuous first derivative at each degeneracy point.
In order to determine the form of the leading logarithmic corrections
we will adopt a low-energy effective field theory (EFT) approach,
applicable near $\myvec{p}_c$.
Here we will discuss the leading logarithmic corrections
near the primary singularities.  For the secondary singularities,
see App.~\ref{secondary} and Refs.~\cite{NNYSPRL,NNYSlong}.

Disregarding any massive states,
we require that the continuum limit of the reduced model is a theory
of relativistic free massless fermions.  This means that there are no massless bosonic
states, and that the massless fermions interact via irrelevant
interactions only, \ie, interactions which vanish
in the continuum limit $a\myvec{p}\to 0$.
These interactions must moreover be consistent with the symmetries
of the underlying reduced model.  The leading logarithmic correction
near a given primary singularity
will arise from a self-energy diagram with two vertices
of the least-irrelevant interaction $\Oirr$ in the EFT
that couples to the corresponding massless fermion state.
Nonanalytic self-energy corrections can arise only if all the
intermediate states are massless.  This is why we can ignore all
massive asymptotic states in this discussion.\footnote{
  Gapped states of the lattice hamiltonian do play a role, however,
  for the secondary singularities.  See App.~\ref{secondary}.
}
Since we deal with an EFT, the scaling dimension $n$ of $\Oirr$
is just its canonical mass dimension, which is integer.
Being irrelevant means that $n>d+1$, or equivalently, since $n$ is integer,
$n\ge d+2$.
The coupling constant of $\Oirr$ can be written as $G a^{n-d-1}$
where $G$ is dimensionless.  The self-energy diagram with two $\Oirr$ vertices
thus gives rise to the generic logarithmic corrections
\begin{subequations}
\label{relzlog}
\begin{eqnarray}
\label{relz1log}
E &=& \pm q\, \Big(1 + c_1 G^2 (aq)^{2(n-d-1)} \log(q^2) \Big) + \cdots \ ,
\hspace{7.8ex} d=1 \ ,
\\
\label{relz3log}
H_{2\times 2} &=& \pm\,\vec\s\cdot\vec{q}\,
\Big(1 + c_3 G^2 (aq)^{2(n-d-1)} \log(q^2) \Big) + \cdots  \ ,
\qquad d=3 \ . \hspace{5ex}
\end{eqnarray}
\end{subequations}
where $\vec{q}=\vec{p}-\vec{p}_c$ and $c_1,c_3$ are numerical constants.
Since $n-d-1\ge 1$, it follows that $\Heff$ has at least
a continuous second derivative at each primary singularity.
In App.~\ref{secondary} we establish that the same is true
at the secondary singularities.

At this point we have established that all the conditions of the NN theorem
are satisfied by $\Heff$, thereby proving the following generalized
no-go theorem:

\begin{quotation}
Consider a reduced model defined on a regular spatial lattice,
with a compact global symmetry $G$ that is not broken spontaneously.
The $G$ generators are thus discrete-valued conserved charges.
Assume also: (1) The hamiltonian has a finite range, and depends
on fermion fields only; (2) The continuum limit is a theory of
relativistic free massless fermions and with no massless bosons;
(3) In any charge sector which supports at least one massless fermion,
one can find a complete set of interpolating fields (as defined in
the introduction) so that the corresponding $\car(\myvec{p})$ is free of zeros.
Then $\Heff(\myvec{p})$ satisfies all the assumptions of the Nielsen-Ninomiya
theorem, and as a result, the massless fermion spectrum
in this charge sector is vector-like.
\end{quotation}

Before we continue, we digress to briefly explain
how the generalized theorem works in the case of a {\em free} hamiltonian.
We require translation invariance, so that the hamiltonian has the general form
$\hat{H} = \sum_{\vec{x},\vec{y}} \sum_{ab}
\j^\dagger_a(\vec{x}) \ch_{ab}(\vec{x}-\vec{y}) \j_b(\vec{y}).$
It can then be shown that, as expected, $\Heff(\vec{p})$ is equal
to $\ch(\vec{p})$, where $\ch(\vec{p})$ is the
Fourier transform of $\ch(\vec{x}-\vec{y})$.
A caveat is that the set of interpolating fields used in the construction
of the retarded and advanced two-point functions must include
all the fermion fields occurring in the hamiltonian.
Omitting some of these fields will result in $\Heff(\vec{p})$ which is
different from $\ch(\vec{p})$.  Moreover, this can lead
to the appearance of zeros in $\car(\vec{p})$,
and thus to spurious poles in $\Heff(\vec{p})$.  For an example
of this phenomenon in the context of the ZZWY model, see App.~\ref{Hpinv}.

In a strict technical sense,
many attempts to construct lattice chiral gauge theories
will not be subject to the generalized no-go theorem as stated above,
if some of its assumptions are not satisfied.
At the technical level, the theory could for example
be defined in euclidean space instead of by a lattice hamiltonian;
or the theory may contain massless scalar fields besides
the fermion fields.  In two dimensions the situation gets
further complicated because four-fermion operators without derivatives
are marginal, as we discuss in Sec.~\ref{nogo2dim} below.
However, one expects that the generalized theorem
is still relevant, because in many cases one can still construct from the
fermion two-point functions of the reduced model at $\o=0$
an object with properties similar to those of $\car(\vec{p})$,
including in particular the analyticity properties if the underlying theory
is local.  Letting this object play the role of $\car(\vec{p})$ can
again give rise to an $\Heff(\vec{p})$ that satisfies all the requirements
of the NN theorem.

Generally speaking, if massless bosons emerge in an SMG phase, their
role in the physics of the continuum limit would have to be understood.
In four dimensions, the presence of
a massless scalar usually signals the spontaneous breaking of
a global symmetry, for which this massless scalar is a Nambu-Goldstone boson.
This is a situation we would like to avoid in the SMG framework.
In two dimensions
continuous global symmetries cannot be broken spontaneously \cite{MW,Coleman},
and long-range order is replaced by quasi-long-range order.
Again, this is a situation we would normally like to avoid.
The reason is simply that
reduced models obtained by turning off the gauge field
in a {\em continuum} chiral gauge theory have a massless spectrum
that consists of fermions only.

In the rest of this section we discuss the implications of the
generalized no-go theorem in four dimensions.
The implications for two-dimensional theories, and for the ZZWY model
in particular, are deferred to Sec.~\ref{nogo2dim} and Sec.~\ref{roadmap}.

Our first comment is that, in four dimensions, the effective theory
that describes the long-distance behavior of a set of massless fermions
is automatically a theory containing irrelevant interactions only.
This is because the interaction with the lowest possible mass dimension,
namely a four-fermion interaction, is irrelevant.

An important exception is the gauge-fixing approach to the construction
of four-dimensional lattice chiral gauge theories, in which one
couples, say, the left-handed fermions to the gauge field, while the
right-handed fermions are spectators.\footnote{%
  The right-handed fermions decouple because of a shift symmetry \cite{GP}.
}
This construction allows for a
Wilson term to remove the doublers at the nonzero corners
of the Brillouin zone. The gauge symmetry is
broken on the lattice, but with gauge fixing and counter terms restoring
Slavnov-Taylor identities,
it is recovered in the continuum limit \cite{nachgt}.
In the corresponding reduced model one can construct an object
that satisfies most of the properties of $\Heff(\vec{p})$.  In particular,
it is hermitian, and analytic away from the (only) degeneracy point
at $\vec{p}=0$, which is a primary singularity.\footnote{%
  The gauge-fixing approach is defined in euclidean space.
  In this context we define the degeneracy points as the points
  where the would-be $\Heff(\vec{p})$ is not analytic,
  and the primary singularities as the points where moreover
  this $\Heff(\vec{p})$ has a zero eigenvalue.
}
However, this would-be $\Heff(\vec{p})$ does not
have a continuous first derivative at the primary singularity
in some channels.  The origin of this behavior
is the presence of a {\em higher-derivative} scalar field in
the reduced model of the gauge fixing approach,
which exhibits a coupling-constant dependent critical exponent.
In each charge sector, for one handedness the primary singularity
is created by the two-point function of a composite operator,
which in turn factorizes as the product of decoupled
scalar and massless fermion two-point functions.
The corresponding sector of $\Heff(\vec{p})$
does not have a continuous first derivative \cite{BGSPRL,BGSlat97b}.

This works as follows.  The group-valued scalar field $\f(x)\in G$
arises in the reduced model of the gauge-fixing approach as the longitudinal
gauge degree of freedom.  Its higher derivative kinetic term
originates from a covariant gauge-fixing term which is part
of the lattice action in this approach.  For $G={\rm U(1)}$, one has
$(\partial_\m A_\m)^2 \to (\bo\f)^2$ when we project $A_\m$ onto its
longitudinal part, $A_\m \to \partial_\m \f$.  The higher-derivative
kinetic term is present in the nonabelian case as well.
The price to pay is an enlarged Hilbert space,
that needs to be projected onto a unitary subspace in the continuum limit.
This teaches us that there might exist
valid dynamical scenarios in which $\Heff(\vec{p})$ will not have
a continuous first derivative at its degeneracy points,
but also that such dynamics has got to be rather non-trivial.
In the gauge-fixing approach the ultimate result is that,
in the continuum limit, the fermion spectrum is free, massless,
and chiral with respect to the (unbroken) symmetries of the reduced model.
In addition, there is a decoupled unphysical sector associated
with the higher-derivative scalar field which, as explained above,
represents the longitudinal gauge degree of freedom.
In the nonabelian case, the unphysical sector
contains the ghost fields as well.

\section{\label{slimit} Decoupling in the strong coupling limit}
In this section we turn to a different, but related, topic.
We consider reduced models in which the fermion fields can be divided
into two sets.  The first set, denote $\c$, includes all the fermion
degrees of freedom that participate in some strong interaction.
The other set, denoted $\x$, includes the remaining fermion degrees of freedom,
that do not directly participate in any strong interaction.
We will then show that in the (uniform) strong-coupling limit the $\x$ and $\c$
sectors decouple.  The relevance of this result is that
the decoupled $\x$ sector is by assumption either free or weakly coupled.
Therefore, the NN theorem applies to the $\x$ sector, and
its spectrum must be vector-like in this limit.

To avoid unnecessary technicalities we will present the main result
while considering a reduced model that depends on fermion fields only,
which are subject to strong interactions only.  We will later comment
on generalizations that broaden the scope of the result.
The total hamiltonian thus has the form\footnote{
Of course, one can alternatively work in a lagrangian formalism.
}
\begin{equation}
\label{totH}
H = H_0(\x,\c) + \Hint(\c;g_1,g_2,g_3,\cdots) \ .
\end{equation}
Here $g_1,g_2,g_3,\ldots$ are the coupling constants of the various
interactions terms, all strong, that couple the $\c$ degrees of freedom.
By assumption, the $\x$ degrees of freedom occur only
in the bilinear part $H_0$ of the hamiltonian,
where they can mix with the $\c$ degrees of freedom.

We now consider the uniform strong-coupling limit defined by writing
\begin{equation}
\label{totHrat}
H = H_0(\j,\c) + \Hint(\c;g_1,g_1\l_2,g_1\l_3,\cdots) \ ,
\end{equation}
where $\l_i=g_i/g_1$, $i=2,3,\ldots$, and then taking the limit
$g_1\to\infty$ while holding all the $\l_i$ fixed.
We first assume that all the interaction terms have the same degree of
homogeneity $n$. We can then rescale the interacting degrees of freedom,
\begin{equation}
\label{rescale}
\c \to g_1^{-1/n} \c \ ,
\end{equation}
and the hamiltonian will take the form
\begin{equation}
\label{totHres}
H = H_0(\j,g_1^{-1/n} \c) + \Hint(\c;1,\l_2,\l_3,\cdots) \ .
\end{equation}
Note that now $g_1$ occurs explicitly only in $H_0$.
Finally taking the limit $g_1\to\infty$ at fixed $\l_i$,
the hamiltonian becomes
\begin{equation}
\label{totHlim}
H = H_0(\x,0) + \Hint(\c;1,\l_2,\l_3,\cdots) \ .
\end{equation}
The $\c$ and $\x$ sets have now decoupled: the $\c$'s occur only in $\Hint$,
while the $\x$'s occur only in $H_0$.  Since $H_0$ is bilinear,
it is subject to the NN theorem, and thus the spectrum of the $\x$'s
must be vector-like in this limit, if $H_0(\x,0)$ satisfies the conditions
of the theorem.  This is the main result of this section.

If the degree of the interaction term with coupling $g_i$ is $n_i$,
with not all $n_i$ equal, we can consider the following
strong-coupling limit.  We choose $g_1$ as the coupling
(or, one of the couplings) whose degree of homogeneity $n_1$
has the smallest value.
After rescaling $\c \to g_1^{-1/n_1} \c$, the hamiltonian becomes
\begin{equation}
\label{Hni}
H_0(\j,g_1^{-1/n_1} \c)
+ \Hint(\c;1,g_1^{1-n_2/n_1}\l_2,g_1^{1-n_3/n_1}\l_3,\cdots) \ ,
\end{equation}
and after taking the limit $g_1\to\infty$ with the $\l_i$'s fixed as before,
we obtain
\begin{equation}
\label{Hlimni}
H = H_0(\x,0) + \Hint(\c;1,\l_2,\l_3,\cdots,0,0,\cdots) \ ,
\end{equation}
where $\l_2,\l_3,\cdots,$ correspond to any additional interaction terms
with the same degree of homogeneity $n_1$, whereas the following zeros
correspond to all other interaction terms, whose degree of homogeneity
is larger than $n_1$.  Once again, the $\x$'s
decouple from the $\c$'s.

These results easily generalize to the case that the reduced model
contains also scalar fields, as well as to the case that the $\x$'s
interact weakly via additional couplings $y_1,y_2,\ldots$, that are kept
fixed (and small) when the uniform strong-coupling limit is taken.
In all of these more general cases, the end result is again that the $\x$'s
decouple from the $\c$'s, and, since the $\x$ sector is either free
or weakly coupled, the $\x$'s are subject to the NN theorem.

A similar result was previously derived in the so-called waveguide model
\cite{wvgd}.  That result is now seen to be a special case of the more general
phenomenon considered here.

Another model where a similar strong-coupling limit was considered
is the Eichten-Preskill (EP) model \cite{EP}, analyzed in detail
in Ref.~\cite{GPR}.  The model has a strong-coupling symmetric (PMS) phase.
In that phase, however, all the fermion degrees of freedom participate
in the strong interaction.  In other words, the $\x$ set is empty.
As for the $\c$ set (which includes all the fermion
degrees of freedom), a strong-coupling expansion
can be used to show that the spectrum in the PMS phase is vector-like,
with the fermion mass tending to infinity
in the strong-coupling limit.\footnote{%
  The PMS phase also has massive composite scalar bound states,
  and the boundary of the PMS phase is defined by the their
  mass-squared going negative, indicating the onset of a phase with
  spontaneous symmetry breaking.
}
The massive Dirac fermions have one chirality component which
is elementary in terms of the lattice fields, while the other
chirality component is composite.  This follows the scenario of
bound-state formation we have discussed in Sec.~\ref{bounds}.

As for the \ZZWY\ model,
the interaction hamiltonian couples only the edge-B fermions, the $\c$'s,
whereas the edge-A fermions, the $\x$'s, occur only in the bilinear hamiltonian.
By the theorem derived in this section, in the strong-coupling limit
the $\c$'s and $\x$'s decouple.  The $\x$ sector becomes free,
and its massless spectrum consists of a single Dirac fermion per
fermion species.  For more details, see App.~\ref{HpSlim}.

An open question regarding the ZZWY model is whether
its phase diagram consists of two phases only (those found in Ref.~\cite{ZZWY}),
or, alternatively, that three (or more) phases may exist.  In the latter case
the SMG phase could be an intermediate phase, which borders
a weak-coupling phase on one side, and a strong-coupling phase
on the other side.  More detailed information on the phase diagram
should help understand to what extent the strong-coupling limit
we studied here is relevant for the properties of the SMG phase.

\section{\label{nogo2dim} The generalized no-go theorem in two dimensions}
In this section we turn to the implications of the generalized
no-go theorem for two-dimensional theories, including in particular
the ZZWY model.

The main difference between four dimensions and two dimensions
is the following.  Assume that the asymptotic states of the reduced model
consist of massless fermions only.  As we pointed out in Sec.~\ref{nogo},
in four dimensions this guarantees that all the interactions
of these fermions will be irrelevant.
The reason is simply that
the lowest-dimension interaction is a four-fermion
operator, whose mass dimension is six (or higher; depending on
whether or not it contains derivatives), which is always irrelevant.

In contrast, in two dimensions a fermion field has mass dimension one half.
Derivative-less four-fermion operators thus have mass dimension two,
hence they represent renormalizable interactions.
In the ZZWY model, the continuous global symmetry of the bilinear part
of the hamiltonian is $\U(2)\times\U(2)$. Taking into account
Fermi statistics, there are three linearly independent,
local four-fermion operators which are built from the ``physical''
edge-A fields only, and are invariant under the full symmetry group.
They can be taken to be $j_R j_R$, $j_L j_L$ and $j_R j_L$,
where\footnote{
  We suppress the Dirac matrices, since for two-dimensional Weyl fields
  they are purely phase factors.
}
$j_L=\sum_{I=1,2} \bx_I \x_I$ and $j_R=\sum_{I=3,4} \bx_I \x_I$.
These operators do not require point splitting, and so in the continuum limit
they indeed turn into derivative-less, renormalizable four-fermion
interactions.  When the interactions of the ZZWY model are turned on,
the global symmetry reduces to the smaller group $\U(1)^2$ (see App.~\ref{syms}).
As a result, more four-fermion operators are allowed by
the symmetry of the full theory.  Now there are altogether six operators made
out of the edge-A fields, which can be taken to be
$(\bx_I \x_I)(\bx_J \x_J)$ with $I\ne J$.

Induced four-fermion operators in two-dimensional theories
have largely been ignored in the SMG literature.
In quantum field theory, ``everything that can happen will happen,''
and thus all these four-fermion operators are expected to be induced.
This includes in particular the SMG phase discussed in Ref.~\cite{CGP},
and the SMG phase of the ZZWY model.

In the face of this situation, what options are available?
The original goal of the SMG paradigm is to construct
lattice chiral gauge theories which are not ``contaminated'' by any other
interactions or massless states not present in the target continuum theory.
In two dimensions, this means that the coefficients of all the induced
four-fermion interactions without derivatives must be tuned to zero by adding
suitable ``counterterms'' to the lattice theory.  Doing so would clearly be
a formidable task, given the large number of the expected four-fermion
interactions and the technical challenge of determining the
induced coupling of every one of them.  Nevertheless,
let us assume that this has been done in a given two-dimensional
reduced model.\footnote{%
  Notice that it is a non-trivial dynamical question whether an SMG phase
  would survive such tuning or not.
}
The continuum limit would then be a theory
of free massless fermions, which is therefore subject to the generalized
no-go theorem of Sec.~\ref{nogo}, just like in the four-dimensional case!
The massless fermion spectrum in the continuum limit
will therefore be vector-like.

In the actual ZZWY model no effort was made to trace the induced
four-fermion interactions, and certainly not to tune them to zero.
It must therefore be assumed that all four-fermion interactions
consistent with the symmetries of the model are present
with $O(1)$ couplings.\footnote{%
  The actual point splittings of the 6-fermion interaction terms
  of the ZZWY model were not specified in Ref.~\cite{ZZWY}.
  Assuming concrete point splittings for definiteness,
  we have checked that (derivative less) four-fermion interactions
  built from the edge-A fields are induced naturally within the
  strong-coupling expansion discussed in Sec.~\ref{slimit}.
}
This raises the next question:  Is it possible to evade the generalized
no-go theorem and build a two-dimensional lattice chiral gauge theory
at the price of inducing additional four-fermion interactions?
Technically, the presence of marginal (equivalently, renormalizable)
interactions in the reduced model invalidates the requirement
that $\Heff(p)$ has a continuous first derivative at all the degeneracy points.
Indeed, if the self-energy correction in Eq.~(\ref{relz1log}) comes from
a marginal operator, we will have $n-d-1=0$.  Hence, while $E(p)$
remains a continuous function of $p$, its derivative is not.
A similar conclusion applies if the reduced model has massless
bosonic states in addition to the massless fermions, which allows for
additional relevant and marginal operators.

Nevertheless, closer scrutiny reveals that doublers may still be present.
Our strategy at this point is to appeal to the simplest form
of the no-go theorem put forward by Karsten and Smit \cite{KS},
and apply it to $\Heff(p)$.  We consider an eigenvalue $E(p)$ of $\Heff(p)$,
assuming only that the dispersion curve $E(p)$ is continuous.
Then, as $p$ traverses the periodic Brillouin zone, every crossing
of $E(p)$ from negative to positive values must be followed by a crossing
in the opposite direction, from positive to negative values.\footnote{%
  In general the dispersion curve $E(p)$ can wrap around the
  Brillouin more than once.  In that case, crossings from both
  negative to positive and from positive to negative values
  can happen at the same value of $p$.  Notice also that
  a discontinuous derivative of the form $E(p) \sim |p|$ is not consistent
  with relativistic invariance.
}

If, at a zero crossing of $E(p)$ from negative to positive
(positive to negative) values, the first derivative is continuous as well,
then this zero crossing is a primary singularity associated
with a RH (LH) massless fermion.  When the first derivative is not continuous
at the crossing, the physical nature of the crossing point may not be
easy to understand, and it requires a detailed investigation within the model
under consideration.   Without such a detailed investigation
it may not be possible to determine whether or not an independent
massless fermion state can be associated with each crossing point.
In App.~\ref{secondary} we present a few examples of weakly coupled
two-dimensional theories with renormalizable four-fermion interactions,
in which the zero eigenvalues of $\Heff(p)$ in the relevant channels
can be identified straightforwardly as massless RH or LH fermions,
and thus the spectrum is vectorlike.

Non-trivial examples
include the strong-coupling symmetric phase of the Smit-Swift model
in two dimensions \cite{BDFS}, and the gauge-fixing approach to the
construction of four-dimensional lattice chiral gauge theories,
already discussed in Sec.~\ref{nogo}.  In the latter case the essential
feature is the presence of a higher-derivative scalar with a $1/(p^2)^2$
propagator, and thus a similar behavior might be found in two dimensions
in the presence of an ordinary massless scalar with a $1/p^2$ propagator.
Of course, in both cases some mechanism must be present to tame
the infrared divergence associated with the scalar field.\footnote{%
  For the gauge-fixing approach, see Refs.~\cite{BGSphase,BGSPRL,BGSlat97b}.
}
We stress that in all of these cases, after turning the gauge field back on
and taking the continuum limit, additional massless fields and/or
additional relevant or marginal interactions will necessarily be present,
on top of the physical fields and interactions of the target
chiral gauge theory.  However, in the case of the gauge-fixing approach
all the additional degrees of freedom are unphysical:
they include the longitudinal degrees of freedom of the gauge field
and ghost fields, that can be shown to decouple from the physical sector
in the usual way (provided that the fermion spectrum is anomaly free),
at least to all orders in perturbation theory.

\section{\label{roadmap} ZZWY model --- the open questions}
We have discussed the conditions for the applicability of the
generalized no-go theorem in Sec.~\ref{nogo},
and the special properties of two-dimensional theories in Sec.~\ref{nogo2dim}.
It should be clear from these discussions that at this point we
have not reached
a final verdict concerning the ZZWY model.  In this section we will
summarize the current situation, and offer a ``road map'' consisting
of the open questions that would have to be sorted out before it can be
firmly established whether or not the SMG phase of the ZZWY model
successfully reproduces the chiral massless spectrum of the 3450 model
in the continuum limit. While this section is tied to the
ZZWY model, the questions we formulate
will need to be considered for any other SMG model as well.

Our starting point is the following observation.
Let us consider the retarded anti-commutator
$R(\myvec{x},t)=R_{\rm elm}(\myvec{x},t)$, where the subscript ``elm''
indicates that the set of lattice fields used in the definition~(\ref{Retc})
consists of the {\em elementary} fields of the model only.
In the notation of Sec.~\ref{bounds} this set is $\{\x_I,\c_I\}$,
where $\x_I$ are the edge-A fields and $\c_I$ the edge-B fields.
We similarly attach the same subscript to the associated correlators
defined by Eqs.~(\ref{Retcx}) --~(\ref{limRetc}) as well as to
the corresponding one-particle hamiltonian
$H_{\rm eff,elm}(p)=\car^{-1}_{\rm elm}(p)$ defined via Eq.~(\ref{Heff}).
The properties of these correlators were not studied in Ref.~\cite{ZZWY}
in sufficient detail, yet the general considerations found
in the condensed matter literature \cite{WYreview,Youetal,YWOX,SYX},
and reviewed in Sec.~\ref{bounds}, strongly suggest that $\car_{\rm elm}(p)$
develops zeros in the SMG phase,
and thus $H_{\rm eff,elm}(p)$ develops poles.

Naively, the presence of poles in $H_{\rm eff,elm}(p)$ appears to be good news,
as it suggests that the NN theorem does not apply and thus the
massless fermion spectrum might yet be chiral in the SMG phase.
However, as we showed in Ref.~\cite{zeros},
if the effective low-energy theory contains poles in (the bilinear part of)
its hamiltonian, this implies the existence of {\em ghost} states
which make the model inconsistent.

The next development came with Ref.~\cite{Youetal}.  This paper discussed
a model where $H_{\rm eff,elm}(p)$ indeed has poles in the SMG phase.
Nevertheless, much like in the EP model \cite{EP},
a strong coupling expansion can be used to prove that there are no
ghost states in the SMG phase \cite{GPR}.
This leads us to the following conjecture,
that we now believe is valid very generally, including in the SMG phase
of the ZZWY model, even though no strong-coupling expansion exists in this case.
The conjecture consists of two parts.  The first part is that
{\em if the underlying theory is local then
the effective low-energy theory cannot contain ghost states
anywhere in the phase diagram.}
The second part is that, as a corollary,
{\em it should always be possible to construct a complete set
of interpolating fields, and thus $\car(p)$ for this set of interpolating fields
has no zeros}.
To recall, a complete set of interpolating fields
was defined in the introduction section by the two requirements
that $\car(p)$ is free of propagator zeros, while the zeros of $\Heff(p)$ are
in one-to-one correspondence with the massless fermion asymptotic states
in the given channel.

According to this conjecture any propagator zeros found in an SMG phase
must be {\em kinematical} zeros.  What this means is that in the SMG phase
there exist additional asymptotic states (apart from those accounted for
by the elementary fermion fields), and the propagator zeros arise because
of missing interpolating fields for these new asymptotic states.
As we proposed in Sec.~\ref{bounds}, the physical mechanism at work is
bound-state formation.  That mechanism was established in the EP model using
its strong-coupling expansion \cite{GPR}; we similarly expect that it can be
established in the model of Ref.~\cite{Youetal}, again using the strong-coupling
expansion.

Addressing the ZZWY model, which lacks a strong-coupling expansion in the
SMG phase, we proposed in Sec.~\ref{bounds} a general procedure for identifying
the bound states that combine with the gapped elementary mirror fermions
to form massive Dirac fermions.
We stress that while these massive Dirac fermions
are expected to decouple in the continuum limit, in order to avoid
propagator zeros one should nevertheless add to the set of
interpolating fields suitable composite fields to generate the
bound-state component of each massive Dirac fermion.\footnote{
  See Sec.~\ref{bounds} for further discussion of scenarios
  for the construction of a complete set of interpolating fields
  in the SMG phase of ZZWY model.
}
It goes without saying that further (numerical) investigations of the
ZZWY model are required in order to establish whether the scenario
of bound-state formation is indeed at work in the SMG phase.  If true,
a complete set of interpolating fields can be constructed in the SMG phase
of the ZZWY model in every charge sector which supports (single)
massless fermion asymptotic states (as discussed in detail in Sec.~\ref{bounds}),
and one could proceed to examine the conditions
for the applicability of the generalized no-go theorem.

The conditions of the generalized no-go theorem were listed in Sec.~\ref{nogo}.
Condition~(1) is satisfied by construction for the ZZWY model:
the model depends on fermion fields only, and has a finite-range hamiltonian.
Next, following the preceding discussion, we expect condition~(3) to be
satisfied because of our conjecture: there exists a complete set
of interpolating fields at every point in the phase diagram
and in every charge sector which admits single-particle massless
fermion states.

The remaining condition of the theorem is condition~(2),
which asserts that the continuum limit is a theory of free massless fermions.
However, as we have discussed in Sec.~\ref{nogo2dim} it is practically certain
that this condition is {\em not} satisfied by the ZZWY model,
because all the renormalizable
four-fermion interactions consistent with the symmetries of the
interacting theory must have been induced in the SMG phase. This
particular concern is, of course, limited to models in two dimensions.

As we then explained in Sec.~\ref{nogo2dim}, at this point there are
two basic alternatives.  The first is to stick to the goal of obtaining
in the continuum limit a chiral gauge theory
(once the gauge field has been turned back on), without any additional
four-fermion interactions.  This would require the modification of the ZZWY model
by new four-fermion counterterms, whose couplings would be tuned so as
to eliminate the renormalizable four-fermion interactions
that will be induced in the ZZWY model in its present form.
Once this -- technically challenging -- task would have been completed,
it is an open question whether or not the SMG phase will survive.
In any case, once all the renormalizable four-fermion interactions
have been tuned away by the counterterms, condition~(2) of the
generalized no-go theorem will also be satisfied.
As a result, the spectrum will be vector-like.

The only alternative would be to opt for a different goal, by allowing the
target continuum chiral gauge theory to depend on the additional,
induced four-fermion interactions.  We believe this is the best one can
hope for in the ZZWY model in its present form,
given that no attempt was made to track
and subtract the induced four-fermion interactions in this model.
As we have further discussed in Sec.~\ref{nogo2dim}, while the generalized
no-go theorem is technically no longer applicable, also in this case
it is far from obvious that the massless spectrum will be chiral.
We also remark that allowing for four-fermion interactions carries with it
the risk of drastically altering the dynamics of the (chiral) gauge theory.
Indeed it is well known that in two dimensions, four-fermion interactions
can by themselves give rise to fermion mass generation without
symmetry breaking \cite{EW}.

Finally, we mention the direct test of the massless fermion spectrum
carried out in Ref.~\cite{CGP} within their attempt to put
on the lattice the same target theory, the
two-dimensional 3--4--5 abelian chiral gauge theory.
In the reduced model, one considers the two-point function of the
conserved current of the global U(1) symmetry to be gauged,
and calculates (numerically, if necessary) its zero-momentum discontinuity.
As we explained in the introduction, the magnitude
of this discontinuity provides a direct test whether or not
the massless fermion spectrum can coincide with the chiral spectrum
of the target 3--4--5 abelian chiral gauge theory.

\section{\label{conc} Conclusion}
The symmetric mass generation, or SMG,
paradigm aims to construct lattice chiral gauge theories
by finding an SMG phase in the reduced model (where the gauge field
has been turned off) in which the to-be-gauged exact global symmetry
is not broken spontaneously, while the massless fermion spectrum is
chiral with respect to that symmetry, and yields the (anomaly-free)
spectrum of the target chiral gauge theory. To accomplish this,
interactions are introduced in the reduced model whose sole purpose is to gap
the unwanted mirror fermions predicted by the Nielsen-Ninomiya theorem.

The SMG literature stresses that a necessary condition for the success
of the SMG approach is that all symmetries (continuous
or discrete) which are anomalous in the target chiral gauge theory
must be broken explicitly in the lattice theory.
But since the fermion spectrum is determined at the level of the reduced model,
this condition must apply already in the reduced model,
even though the gauge field has been turned off.  In other words,
this condition translates into the requirement that all the
exact (global) symmetries of the reduced model in the relevant SMG phase
must not have an anomaly in the target chiral gauge theory.\footnote{%
  See, for example, Ref.~\cite{WYreview} and references therein.
}
We will refer to this requirement as the SMG anomaly paradigm.

While it seems easy to motivate the SMG anomaly paradigm
at an intuitive level, it has remained at odds with numerous lattice studies.
The paradigm goes back to the Eichten-Preskill model \cite{EP}.  However,
it was shown long ago that the SMG phase of the Eichten-Preskill model
(known in the relevant literature as a PMS phase)
fails to support a chiral massless spectrum \cite{GPR}
for essentially the same dynamical reasons as the corresponding phase
of the Smit-Swift model \cite{BD,BDJJNS,GPS,largeWY},
even though the Eichten-Preskill model adhered to the
SMG anomaly paradigm, whereas the Smit-Swift model did not.
The Nielsen-Ninomiya theorem, and its generalization discussed
in this paper, are also oblivious to the SMG anomaly paradigm.
It is thus a logical possibility that the underlying reason
for the potential failure of the SMG approach is the remarkably wide scope
of the Nielsen-Ninomiya theorem, as reflected by
the generalized no-go theorem, while the SMG anomaly paradigm does
not play an important role.

The original Nielsen-Ninomiya theorem does not apply in an SMG phase,
because it is a theorem about free lattice theories.  However,
the generalization of this theorem to the case that interactions
of arbitrary strength are present applies in principle everywhere
in the phase diagram of any reduced model, including in an SMG phase.
Physically, what allows for the generalized theorem is the requirement
that the continuum limit of the reduced model will be a theory of
free massless fermions.  Of course, for the SMG paradigm to succeed,
the spectrum of these massless fermions must be chiral.
These requirements, if satisfied, ensure that after the gauge field
is turned on one recovers a chiral gauge theory without any additional
``contaminating'' interactions.
But, basically because the continuum limit of the reduced model
has to be a free theory, starting from the fermion
two-point functions of the reduced model it is often possible to construct
a one-particle lattice hamiltonian $\Heff$ that will satisfy
all the conditions of the Nielsen-Ninomiya theorem.
Once the Nielsen-Ninomiya theorem applies, the spectrum must be vector-like.

In this paper, the required properties of $\Heff$ were established
in a fairly specific setting: a lattice hamiltonian defined on
a spatial lattice, which depends on fermion fields only,
and has a finite range.  However, experience teaches us that
objects with essentially the same properties as this paper's $\Heff$
can be constructed in a much more general setting.  This includes
lattice models formulated within the euclidean path-integral framework,
where two-point functions with similar features as $\car(\vec{p})$
can be constructed for $\o=0$, including similar analyticity properties
if the underlying theory is local.  Whenever such a construction is possible,
the Nielsen-Ninomiya theorem eventually applies.

The starting point of this paper was the role of propagator zeros
in an SMG phase, previously discussed in Refs.~\cite{zeros,Youetal}.
Our renewed examination of this issue strongly suggests that a complete set
of interpolating fields whose two-point functions are free of
propagator zeros should exist everywhere in the phase diagram
of any local reduced model built along the lines of the SMG paradigm
(see Sec.~\ref{roadmap} for a summary).
While finding the desired complete set in any particular model might require
a trial-and-error process, we stress that the inability to build
such a complete set would likely imply that the propagator zeros represent
``irremovable'' ghost states.  This situation is not only highly unlikely
given that the underlying theory is local \cite{Youetal}, but moreover,
if realized, would render the theory inconsistent \cite{zeros}.

As we have extensively discussed in this paper,
a key assumption of the generalized no-go theorem
is that the continuum limit of the reduced model is a theory
of free massless fermions.  This requirement leads to a major difference
between four-dimensional and two-dimensional theories.
In four dimensions, once the massless spectrum consists of fermions only,
the continuum limit is automatically a free theory,
because all multi-fermion interactions are irrelevant in four dimensions.
This makes it easier to satisfy the conditions of the generalized no-go theorem.
In contrast, four-fermion interactions without derivatives
are renormalizable in two dimensions.
This not only makes the situation in two dimensions much more complicated
than in four dimensions, it also significantly limits any lessons that
might be drawn from two-dimensional models for the physically interesting case
of four dimensions.

\vspace{3ex}
\noindent
{\bf Acknowledgments.\ }
We thank Yi-Zhuang You and Cenke Xu for fruitful discussions.
MG and YS both benefited from the workshop on Symmetric
Mass Generation held in May 2024 at the Simons Center
for Geometry and Physics at SUNY Stony Brook.
MG also thanks the Kavli Institute of Theoretical Physics at UC Santa Barbara,
where part of this work was carried out during the program
``What is Particle Theory,'' for hospitality.
This material is based upon work supported by the U.S. Department of
Energy, Office of Science, Office of High Energy Physics,
Office of Basic Energy Sciences Energy Frontier Research Centers program
under Award Number DE-SC0013682 (MG).
YS is supported by the Israel Science Foundation under grant no.~1429/21.

\appendix
\section{\label{HZZWY} The \ZZWY\ hamiltonian}
In this appendix we provide some technical details about the hamiltonian
of the \ZZWY\ model \cite{ZZWY}.  In App.~\ref{Hpfree} we discuss
the bilinear part of the hamiltonian $H_0$ and its spectrum.
In App.~\ref{Hpinv} we discuss the inverse of the free hamiltonian,
and give another example of a propagator zero.
In App.~\ref{HpSlim} we
discuss the effect of the decoupling of the edge-A and edge-B fermions
in the strong-coupling limit of Sec.~\ref{slimit}.
In App.~\ref{syms} we briefly discuss
the interaction hamiltonian $\Hint$ and its symmetries.

\subsection{\label{Hpfree} The bilinear hamiltonian}
Here we discuss the bilinear part of the hamiltonian, defined in Eq.~(\ref{H0}).
For $I=1,2$, namely for the fermion species with charges 3 and 4 under the
U(1) symmetry to be gauged, the hamiltonian matrix is $\ch(\t_1,t_2)$;
whereas for $I=3,4$, or the fermion species with charges 5 and 0,
it is $\ch(\t_1^*,t_2)$.  The actual values of
the parameters in the \ZZWY\ model are
\begin{eqnarray}
\label{t1t2}
\t_1 &=& t_1 e^{i\frac{\p}{4}}  \ =\ e^{i\frac{\p}{4}} \ ,
\\
t_2 &=& 1/2 \ .
\nonumber
\end{eqnarray}
Here we will focus on the bilinear hamiltonian for one of the first two species,
$\hH = \bj\ch(\t_1,t_2)\j$, omitting the index $I$.

The fermion fields reside on the lattice shown in \Fig{3450latt},
which constitutes two interconnected one-dimensional chains.
The unit cell of this lattice is a $2\times 1$ rectangle in units of
the lattice spacing $a$.  The (one dimensional) momentum $p$ thus lives
in a Brillouin zone $0\le p \le 2\pi/(2a) = \p/a$.
Henceforth we will set $a=1$.

The unit cell has four independent degrees of freedom, which are
associated with the four sublattices indicated by the numbers inside
the circles in \Fig{3450latt}, which start at the upper-left corner
of the unit cell and go clockwise.  Consistent with Fig.~1(c) of Ref.~\cite{ZZWY},
we will also refer to the upper chain in our figure as edge~A,
and to the lower chain as edge~B.  Hence edge~A consists of
sublattices~1 and~2, while edge~B consists of sublattices~3 and~4.

Labeling the degrees of freedom of the four sublattices as $\j_k$, $k=1,2,3,4$,
one has\footnote{%
  We arbitrarily place sublattices~1 and~4 on the even sites,
  and thus sublattices~2 and~3 on the odd sites.
  We thank YiZhuang You for some clarifications about the structure
  of the lattice.
}
\begin{eqnarray}
\label{H1alt}
\hH &=& \sum_x t_1\Bigl(e^{i\p/4}\j^\dagger_1(2x)(\j_2(2x+1) + \j_2(2x-1)) \\
&&\qquad +\ e^{-i\p/4}\j^\dagger_4(x)(\j_3(2x+1) + \j_3(2x-1)) \nonumber\\
&&\rule{0ex}{3ex}\qquad +\ e^{-i\p/4}\j^\dagger_1(2x)\j_4(2x)
+ e^{i\p/4}\j^\dagger_2(2x+1)\j_3(2x+1) + \hc\Bigr) \nonumber\\
&& +\sum_x t_2\Bigl(\j^\dagger_1(2x)(\j_3(2x-1)-\j_3(2x+1)) \nonumber\\
&&\qquad +\ \j^\dagger_4(2x)(\j_2(2x-1)-\j_2(2x+1)) + \hc\Bigr)\ .
\nonumber
\end{eqnarray}
In this expression we have kept the parameters $t_1,t_2$ free,
to make it easier to trace the origin of the various terms.
In terms of the degrees of freedom associated with the four sublattices,
the hamiltonian matrix in momentum space is
\begin{equation}
\label{Hp}
\ch =
\left(\begin{array}{cccc}
  0 & 2t_1 e^{i\frac{\p}{4}} \cos(p) & -2it_2\sin(p) & t_1 e^{-i\frac{\p}{4}} \\
  2t_1 e^{-i\frac{\p}{4}} \cos(p) & 0 & t_1 e^{i\frac{\p}{4}} & 2it_2\sin(p) \\
  2it_2\sin(p) & t_1 e^{-i\frac{\p}{4}}  & 0 & 2t_1 e^{i\frac{\p}{4}} \cos(p) \\
  t_1 e^{i\frac{\p}{4}} & -2it_2\sin(p) & 2t_1 e^{-i\frac{\p}{4}} \cos(p) & 0
\end{array}\right) \ . \hspace{5ex}
\end{equation}
The spectrum of this hamiltonian is shown in \Fig{3450spec},
using the values of the parameters from Eq.~(\ref{t1t2}).
This reproduces Fig.~1(b) of Ref.~\cite{ZZWY}.

\begin{figure}[t]
\begin{center}
\includegraphics*[width=8cm]{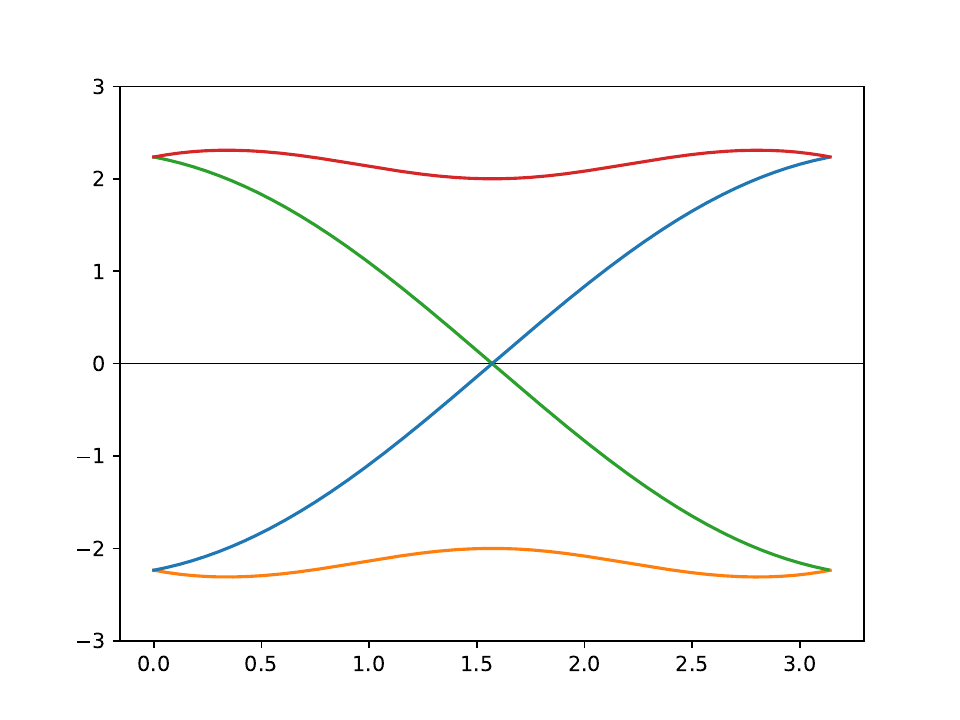}
\end{center}
\vspace*{-5ex}
\begin{quotation}
\floatcaption{3450spec}%
{Spectrum of the bilinear hamiltonian $\hH$,
reproducing Fig.~1(b) of Ref.~\cite{ZZWY}.
The horizontal axis runs from 0 to $\p$.
The green (blue) branch corresponds to the LH (RH) chiral mode
supported mainly on edge~A (edge~B).  The plot shows that
the spectrum has the correct mod $\p$ periodicity.
Also seen is that the eigenvalues spectrum forms a single smooth curve
that winds around the Brillouin zone four times.
}
\end{quotation}
\vspace*{-4ex}
\end{figure}

As can be seen in \Fig{3450spec}, the spectrum consists of two massless
and two massive branches,\footnote{%
  Notice that the spectrum consists of a single continuous curve
  that wraps around the Brillouin zone four times.
}
and the critical momentum where the energy
of the two chiral modes vanishes is $p=\frac{\p}{2}$.
Expanding to quadratic order in the relative momentum $k=p-p_c=p-\frac{\p}{2}$,
and solving the eigenvalue equation, we find that the eigenvectors
of the two chiral modes are (in an arbitrary normalization)
\begin{equation}
\label{vpm}
v_+ = \left(\begin{array}{c}
  e^{-i\frac{\p}{4}} \frac{k}{8} \\
\rule{0ex}{2.5ex}
  -\frac{ik}{8} \\
\rule{0ex}{3ex}
  e^{-i\frac{3\p}{4}} \\
\rule{0ex}{2.5ex}
  1
\end{array} \right) , \hspace{5ex}
v_- = \left(\begin{array}{c}
  e^{-i\frac{3\p}{4}} \\
\rule{0ex}{2.5ex}
  -1 \\
\rule{0ex}{3ex}
  e^{-i\frac{\p}{4}} \frac{k}{8} \\
\rule{0ex}{2.5ex}
  \frac{ik}{8}
\end{array} \right) ,
\end{equation}
while the eigenvectors of the two massive modes are
\begin{equation}
\label{upm}
u_+ = \left(\begin{array}{c}
  e^{-i\frac{\p}{4}} (1 + k + \frac{k^2}{2}) \\
\rule{0ex}{2.5ex}
  i(1 + k + \frac{k^2}{2}) \\
\rule{0ex}{3ex}
  e^{i\frac{\p}{4}} \\
\rule{0ex}{2.5ex}
  1
\end{array} \right) , \hspace{5ex}
u_- = \left(\begin{array}{c}
  e^{i\frac{\p}{4}} \\
\rule{0ex}{2.5ex}
  -1 \\
\rule{0ex}{3ex}
  e^{-i\frac{\p}{4}} (1 + k + \frac{k^2}{2}) \\
\rule{0ex}{2.5ex}
  -i(1 + k + \frac{k^2}{2})
\end{array} \right) .
\end{equation}
The energies are
\begin{subequations}
\label{Epm}
\begin{eqnarray}
\label{Epma}
E(v_\pm) &=& \pm 2k \ ,
\\
\label{Epmb}
E(u_\pm) &=& \pm (2 + \mbox{$\frac{k^2}{2}$}) \ .
\end{eqnarray}
\end{subequations}
For the chiral mode $v_+$, the relative momentum $k$ and energy $E$
have the same sign, making it the RH mode by definition.
For $v_-$ they have opposite signs, and thus it is the LH mode.
As can be seen in Eq.~(\ref{vpm}),
the RH mode $v_+$ is supported mainly on sublattices~3 and~4,
that is, on edge~B, while the LH mode $v_-$ is supported mainly
on sublattices~1 and~2, hence on edge~A.\footnote{%
  For any $k\ne 0$, the support of the chiral mode
  has a small component on the opposite edge.
}

Appealing to the tensor-product matrix notation introduced in App.~\ref{Hpinv}
below, the matrix $[\s_1\otimes\s_3]$ anti-commutes with $\ch(p)$.
Hence, the application of $[\s_1\otimes\s_3]$
to an eigenvector with eigenvalue $E$ gives rise to an eigenvector
with eigenvalue $-E$, as can be verified using the explicit form
of the eigenvectors in Eqs.~(\ref{vpm}) and~(\ref{upm}).
Notice that the factor of $\s_1$ in this tensor product interchanges
the two edges, consistent with the fact that the RH and LH modes
live on opposite edges.

Turning to the last two species in Eq.~(\ref{H0}), it is straightforward to check
that if $v_+(p)$ is an eigenmode of $\ch(\t_1,t_2)$ with
relative momentum $k=p-p_c$, then $v_+^*(p)$ is an eigenmode
of $\ch(\t_1^*,t_2)$ with the same energy $E(p)$ but with an opposite
relative momentum.  The behavior of $v_-(p)$ is similar.
It follows that the roles of the RH and LH modes are interchanged
for $\ch(\t_1^*,t_2)$.

Recall that in Ref.~\cite{ZZWY} the chiral modes on edge~A of the four
fermion species are identified with the fields of the target 3450 model,
and the goal is to gap the mirror modes on edge~B.
It follows that the chiral modes with charges~3 and~4
of the target 3450 model are LH, while the chiral modes with charges~5 and~0
are RH.

\subsection{\label{Hpinv} Inverse of the bilinear hamiltonian and propagator zeros}
Here we construct the inverse of the momentum-space hamiltonian matrix
$\ch(p)$, Eq.~(\ref{Hp}), and examine its behavior near the primary singularity
$p_c=\p/2$.  This allows us to give an example of a propagator zero,
obtained by projecting $\ch^{-1}(p)$ on the fermion fields of one edge only.

We begin by introducing a convenient representation of $\ch(p)$
in terms of tensor products of Pauli matrices,
\begin{equation}
\label{Hprodval}
\ch(p) = -2 \cos(p) [I\otimes\ts_2] + [\s_1\otimes\ts_1]
         + \sin(p) [\s_2\otimes\s_3] \ ,
\end{equation}
where we have used the values of $\t_1$ and $t_2$ in Eq.~(\ref{t1t2}).
The tensor products correspond to the splitting of $\ch(p)$
into $2\times 2$ blocks.
In each tensor product, the first factor corresponds to the block structure,
and as a basis for it we take the usual Pauli matrices
together with the identity matrix $I$.  The second factor
corresponds to the internal structure of the blocks,
and as a basis we use a rotated set of Pauli matrices,
again together with the identity matrix.  The rotated matrices are
\begin{eqnarray}
\label{rotP}
\ts_1 &=& e^{-\frac{i\p}{8}\s_3} \s_1 e^{+\frac{i\p}{8}\s_3}
= 2^{-1/2} (\s_1+\s_2) \ , \\
\ts_2 &=& e^{-\frac{i\p}{8}\s_3} \s_2 e^{+\frac{i\p}{8}\s_3}
= 2^{-1/2} (\s_2-\s_1) \ , \nonumber\\
\ts_3 &=& \s_3 \ .
\nonumber
\end{eqnarray}

We start by noting that
\begin{equation}
\label{Hsq}
\ch^2(p) = 2 + 3\cos^2(p) + 2\sin(p) [\s_3\otimes\ts_2] \ .
\end{equation}
The tensor-product matrix $[\s_3\otimes\ts_2]$ commutes with $\ch(p)$.
Using Eq.~(\ref{Hsq}) it is straightforward to check that
\begin{equation}
\label{Hinv}
\ch^{-1}(p) = \ch(p) \Big(2 + 3\cos^2(p) - 2\sin(p) [\s_3\otimes\ts_2] \Big)
\Big(16 \cos^2(p) + 9\cos^4(p)\Big)^{-1} .  \hspace{3ex}
\end{equation}
Again considering small $k=p-\p/2$ near the primary singularity $p_c=\p/2$,
and introducing projectors
$P_\pm = \half (1 \pm [\s_3\otimes\ts_2])$, we find
\begin{equation}
\label{Hinvpc}
\ch^{-1}(k) = P_-\, [I\otimes\ts_2]\, \frac{1}{2k} +
P_+ [\s_1\otimes\ts_1] \frac{1}{2} + \cdots \ ,
\end{equation}
which is consistent with the eigenvalue spectrum~(\ref{Epm}).
Notice that the first term on the right-hand side captures
the two massless chiral modes of opposite chiralities, Eq.~(\ref{Epma}).

We next turn to another example of a propagator zero.
We recall that in the free case $\car(p)=\ch^{-1}(p)$ (see Sec.~\ref{nogo}).
$\ch^{-1}(p)$ has a LH pole on edge~A, and a RH pole on edge~B.
We thus expect that a propagator zero will appear if we project $\ch^{-1}(p)$
onto the fields of one edge only, since in this case the pole originating
from the other edge will be missing.

The projectors on the edge-A and edge-B
fields are $P_{A,B} = \half(1\pm [\s_3\otimes I])$.  Let us check
what happens if we project the propagator on (say) the edge-A fields.
First, if we sandwich $\ch(p)$ itself between $P_A$ projectors the result is
\begin{equation}
\label{HprojA}
P_A \ch(p) P_A = -2 \cos(p) P_A [I\otimes\ts_2] P_A \ .
\end{equation}
We next use that $P_{A,B}$ commute with $P_\pm$, as well as
with $[I\otimes\ts_2]$.  Again considering small $k=p-\p/2$ we find
\begin{equation}
\label{HpPa}
P_A \ch^{-1}(k) P_A = P_A [I\otimes\ts_2]
\left( P_-\, \frac{1}{2k} + P_+\, \frac{k}{2} \right) + \cdots \ .
\end{equation}
In order to understand the physical content of this result,
notice that within the matrix block form we are using,
the two $P_A$ projectors on the left-hand side of Eq.~(\ref{HpPa})
pick out the upper-left 2-by-2 block of $\ch^{-1}$,
which is the diagonal block with $\s_3=+1$.  Making the substitution $\s_3=+1$,
the $4\times 4$ projectors $P_\pm$ reduce to $2\times 2$ projectors
$\tP_{2\pm} = \half(1 \pm \ts_2)$.  Hence
\begin{eqnarray}
\label{HpPa2by2}
(P_A \ch^{-1}(k) P_A)_{2\times 2} &=&
\ts_2 \left( \tP_{2-}\, \frac{1}{2k} + \tP_{2+}\, \frac{k}{2} \right) + \cdots
\\
&=& \left( -\tP_{2-}\, \frac{1}{2k} + \tP_{2+}\, \frac{k}{2} \right) + \cdots \ .
\nonumber
\end{eqnarray}
While the first term on the right-hand side is the familiar LH pole
of edge~A, the missing RH pole of edge~B got replaced by a propagator zero.
(This is a RH zero in the sense that the corresponding eigenvalue
of the propagator has the same sign as $k$.)
The eigenvalues of the $2\times 2$ matrix~(\ref{HpPa2by2})
form a single curve with the following properties:
the curve covers the Brillouin zone twice,
it is continuous everywhere except at the LH pole,
and it has no additional zeros except for the RH zero
we have found at $p=\p/2$.

We show the eigenvalues of the complete $\ch^{-1}(p)$ in the left panel of
\Fig{Hinvspec}.  The poles of the LH and RH chiral modes, which are supported
on edge~A and edge~B respectively, are clearly seen.  In the right panel,
we show the eigenvalues of the upper-left 2-by-2 block of $\ch^{-1}(p)$.
This corresponds to a two-point function in which we retain only
the edge-A degrees of freedom.  The LH pole, which is supported mainly
on edge~A, is affected only in a minor way.  But the RH pole, which is
supported mainly on the missing edge-B degrees of freedom, is absent.
In its place, there is now a -- kinematical -- propagator zero.

\begin{figure}[t]
\begin{center}
\includegraphics*[width=7.5cm]{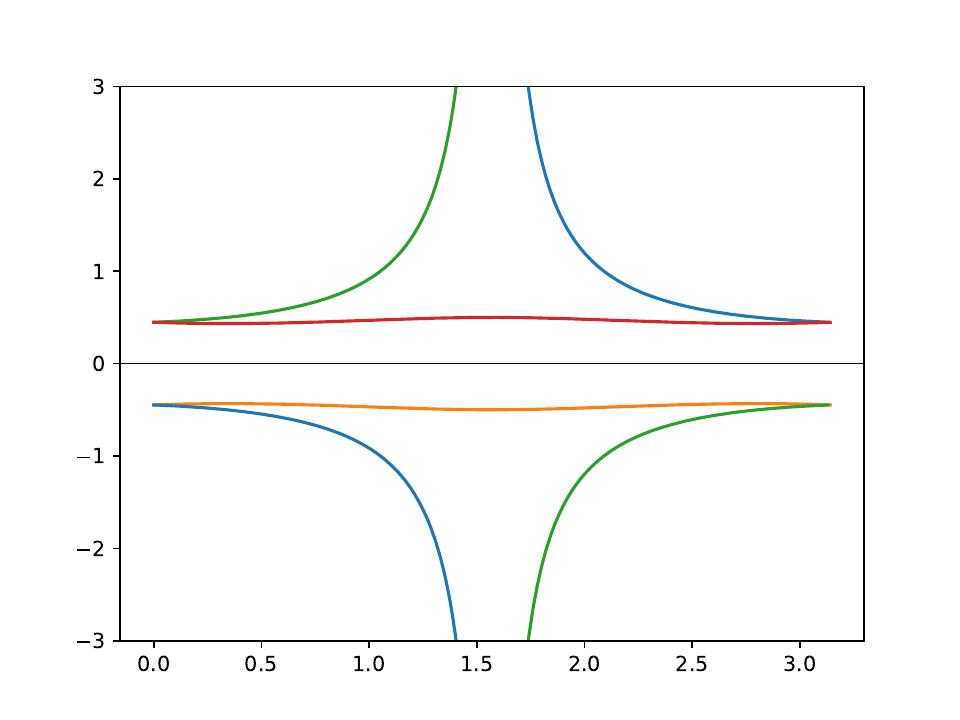}
\includegraphics*[width=7.5cm]{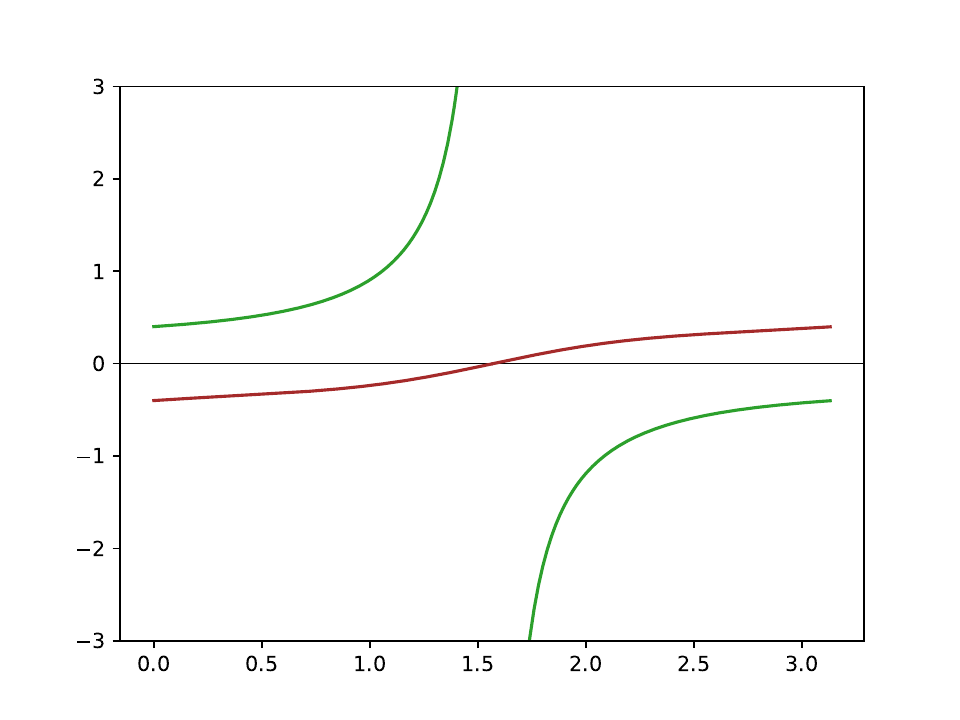}
\end{center}
\begin{quotation}
\floatcaption{Hinvspec}%
{Eigenvalues of $\ch^{-1}(p)$.
Left: eigenvalues of the full $\ch^{-1}(p)$.
Colors match those in \Fig{3450spec}.  In particular, the poles of the LH (RH)
chiral modes are shown in green (blue).  Right: eigenvalues of the
upper-left 2-by-2 block of $\ch^{-1}(p)$.  While the LH pole (green)
supported near edge~A is reproduced, the RH pole got replaced by
a propagator zero.
}
\end{quotation}
\vspace*{-4ex}
\end{figure}

Finally, to avoid confusion we would like to stress the difference between
what we have discussed here, and the situation described in App.~\ref{HpSlim}
below. Here we are interested in the effect of applying a projection
to the inverse, $P_A \ch^{-1}(p) P_A$, and, in particular, the propagator zero
that this projection generates.  By contrast, in App.~\ref{HpSlim} the
question is what is the physical effect of applying a projection
to the hamiltonian itself, $P_A \ch(p) P_A$ (already given in Eq.~(\ref{HprojA})),
as we will be considering a strong-coupling limit that turns
the edge-A fermions into a free theory, decoupled from the edge-B fermions.

The appearance of a zero in the projected inverse $P_A \ch^{-1}(p) P_A$
is yet another example of a general phenomenon we have discussed
in this paper: if our set of interpolating fields is under-complete,
and the poles of some components of the fermion asymptotic states are missing
from its two-point function, this is bound to lead to zeros in the same
two-point function.  Specifically, the case we have considered here
corresponds to constructing\footnote{
  In the free limit where all the interactions are turned off, of course.
}
the (retarded) two-point function of the edge-A fields only,
while leaving out the edge-B fields, even though the fields of both edges
are coupled in the bilinear hamiltonian.

Of course, when we deal with
a bilinear hamiltonian there is no difficulty to identify an appropriate
set of interpolating fields.  As already pointed out in Sec.~\ref{nogo},
in this case $\Heff(\vec{p})$ becomes equal to the momentum-space
hamiltonian matrix $\ch(\vec{p})$, provided that our set of
interpolating fields includes all the fields appearing in
the hamiltonian.  Assuming that the original free hamiltonian
is short ranged, $\ch(\vec{p})$ is an analytic function of $\vec{p}$,
and thus so is $\Heff(\vec{p})$.  This is true regardless of
the precise relation between the lattice fermion fields and the
massless fermion states of the hamiltonian.  In particular,
it is possible that a particular lattice field generates more than one
massless fermion state, as in the case of naive fermions; or that
certain linear combinations of the lattice fields do not generate
any massless states near a particular primary singularity,
as in the case of the massive modes of the ZZWY hamiltonian (Eq.~(\ref{Epmb})).

By contrast, in the presence of potentially strong
interactions, such as in an SMG phase, the presence of propagator zeros in
the two-point function of all the elementary fermion fields may signal
that in this phase the fermion asymptotic states contain new degrees of freedom
which can be generated by composite fields only.

\subsection{\label{HpSlim} Massless spectrum in the strong-coupling limit}
Following the theorem of Sec.~\ref{slimit}, when the coupling constants
of the interaction hamiltonian tend to infinity simultaneously,
the edge-B fermions decouple from the edge-A fermions.
For each of the four species, the edge-A fermions then form
a decoupled free theory.  Since edge~A consists of sublattices~1 and~2,
it follows that their (free) hamiltonian is the upper-left $2\times 2$ block
of the full hamiltonian matrix~(\ref{Hp}), given in Eq.~(\ref{HprojA}).
The spectrum of this ``left-over'' hamiltonian
is shown in \Fig{H3450dbl}.  As expected, it shows doubling.

A comparison of \Fig{H3450dbl} with \Fig{3450spec}
reveals some interesting facts.  First, the critical momentum remains at
$p_c=\p/2$.  Also, at a qualitative level, the effect of the
strong coupling limit is to eliminate the two massive branches from
the spectrum of the bilinear hamiltonian, while the two massless branches
remain qualitatively the same as before.

As for the edge-B degrees of freedom of the four fermion species,
they form a single strongly-coupled system,
for which no strong-coupling expansion is available.
As mentioned in Sec.~\ref{bounds}, this is because of the presence of
hopping terms in the interactions (see also App.~\ref{syms}).

\begin{figure}[t]
\begin{center}
\includegraphics*[width=8cm]{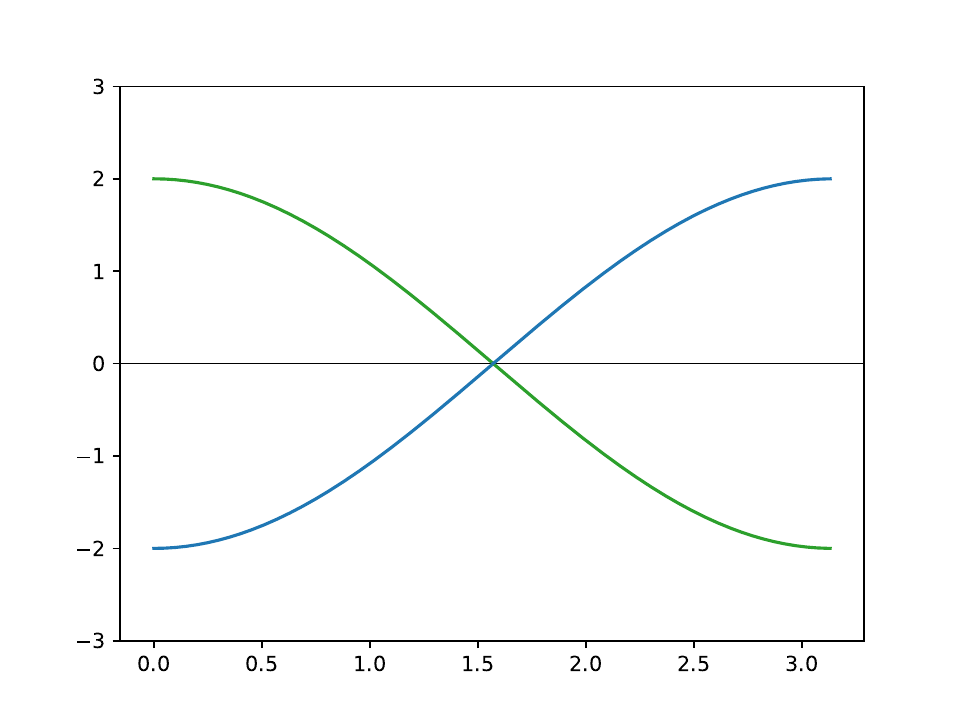}
\end{center}
\begin{quotation}
\floatcaption{H3450dbl}%
{Spectrum of the ``left-over'' hamiltonian---the upper-left $2\times 2$ block
of the hamiltonian matrix~(\ref{Hp}).}
\end{quotation}
\vspace*{0ex}
\end{figure}

\begin{boldmath}
\subsection{\label{syms} $\Hint$ and symmetries}
\end{boldmath}
Here we briefly review the algebraic structure underlying
the construction of the interaction hamiltonian $\Hint$ of the \ZZWY\ model
and its symmetries.  Both the operator content and the symmetries
of $\Hint$ are encoded in the following set of orthogonal vectors,
\begin{subequations}
\label{qell}
\begin{eqnarray}
\label{qella}
\tq_1 &=& (3,4;5,0) \ ,
\\
\label{qellb}
\tq_2 &=& (4,-3;0,-5) \ ,
\\
\label{qellc}
\tl_1 &=& (1,-2;1,2) \ ,
\\
\label{qelld}
\tl_2 &=& (2,1;-2,1) \ .
\end{eqnarray}
\end{subequations}
This set plays a key role in the formal arguments for the decoupling
of the mirrors in the SMG phase, which we will not repeat here.
For the full details, see Refs.~\cite{WWPRB,DMW,ZZWY}.

Consider first the two $\tq$ vectors.  The first vector, $\tq_1$,
is recognized as the set of charge assignments of the four fermion species
under the U(1) symmetry to be gauged.  Its first two entries provide
the charges of the LH fields of the target theory, and its last two entries
provide the charges of the RH fields.  The second vector, $\tq_2$,
defines a linearly independent
set of charge assignments, which is also anomaly free.
By construction, the two U(1) transformations defined by
both $\tq_1$ and $\tq_2$ will be respected by $\Hint$.

The $\tl$ vectors encode the two 6-fermion operators introduced
in $\Hint$, which are
consistent with the U(1) symmetries associated with the two $\tq$ vectors.
Recalling that the interaction hamiltonian involves only the edge-B fermions,
$\tl_1$ corresponds to a 6-fermion operator with the schematic structure
$\c_1(\c_2^\dagger)^2\c_3\c_4^2$, while $\tl_2$
corresponds to $\c_1^2\c_2(\c_3^\dagger)^2\c_4$.
The orthogonality of the four vectors in Eq.~(\ref{qell}) implies
that these 6-fermion operators are invariant under the two U(1) symmetries
associated with the $\tq$ vectors.
Since the elementary fermions have a single degree of freedom
per lattice site, same-field products (for example, $\c_1^2$)
must be point-split.  The corresponding hopping terms
turn into derivatives in the (classical) continuum limit.
We note that Refs.~\cite{WWPRB,DMW,ZZWY} did not discuss the four-fermion
operators consistent with the same U(1) symmetries (see Sec.~\ref{nogo2dim}).

\vspace{0.8cm}

\section{\label{Rp} Properties of two-point functions}
In this appendix we establish some properties of the two-point functions
used in the proof of the generalized no-go theorem in Sec.~\ref{nogo}.
In App.~\ref{Rpdagger} we establish the hermiticity of the relevant
two-point functions for $\o\to 0$.  In App.~\ref{Rpsmooth} we prove
under certain assumptions that $\car(\myvec{p})$
is infinitely differentiable except at the degeneracy points.
In App.~\ref{Rpwedge} we prove the stronger result
that $\car(\myvec{p})$ is analytic except at the degeneracy points.
We recall that $\ca(\myvec{p})=\car(\myvec{p})$
away from degeneracy points (see Sec.~\ref{nogo}).
Finally, in App.~\ref{secondary} we explain the role
of secondary singularities.

\subsection{\label{Rpdagger} Hermiticity}
Here we establish the hermiticity properties of the retarded anti-commutator.
Specifically, we will consider the Fourier transform
of one of the two terms that make it up (see Eqs.~(\ref{Retc}) --~(\ref{limRetc})),
\begin{equation}
\label{Spom}
S_{ab}(\myvec{p},\o) = i\int_0^\infty dt\, e^{i\o t}
\sum_{\myvec{x}} e^{ - i\myvec{p}\cdot\myvec{x}}
\sbra{0}\J_a(\myvec{x},t)\,\J^\dagger_b(\myvec{0},0)\sket{0} ,
\end{equation}
and prove that it is hermitian for $\Im\o\to 0$.
The proof for the other term works in the same way,
as it does for the advanced function.  As we will see below
the field $\J_a(\myvec{x},t)$ can be both elementary or composite.

We begin by noting that
\begin{equation}
\label{psitx}
\J_a(\myvec{x},t) = e^{i(Ht-\myvec{P}\cdot\myvec{x})} \J_a(\myvec{0},0)
e^{-i(Ht-\myvec{P}\cdot\myvec{x})} \ ,
\end{equation}
hence
\begin{equation}
\label{statenp}
\sbra{0}\J_a(\myvec{x},t)\sket{\myvec{p},n}
= e^{-i(E_n(\myvec{p})t - \myvec{p}\cdot\myvec{x})}
\sbra{0}\J_a(\myvec{0},0)\sket{\myvec{p},n}
\equiv e^{-i(E_n(\myvec{p})t - \myvec{p}\cdot\myvec{x})} v_a(\myvec{p},n) \ .
\end{equation}
Here and below, the index $n$ labels all the states with momentum $\myvec{p}$.
We next introduce a complete set of intermediate states
and perform the $\myvec{x}$ summation in Eq.~(\ref{Spom}), which
projects out the intermediate states with momentum $\myvec{p}$,
\begin{equation}
\label{sumxp}
\sum_{\myvec{x}} e^{ - i\myvec{p}\cdot\myvec{x}}
\sbra{0}\J_a(\myvec{x},t)\,\J^\dagger_b(\myvec{0},0)\sket{0}
= \sum_n e^{-iE_n(\myvec{p})t} v_a(\myvec{p},n) v_b^*(\myvec{p},n) \ .
\end{equation}
The matrix elements product $v_a(\myvec{p},n) v_b^*(\myvec{p},n)$ is
manifestly hermitian.  It remains to perform the time integral.
Assuming that $\o$ has a positive imaginary part, for a particular
intermediate state this integral is
\begin{equation}
\label{tint}
i\int_0^\infty dt\, e^{i(\o-E_n(\myvec{p}))t}
= -\frac{1}{\o-E_n(\myvec{p})} \ .
\end{equation}
Finally sending $\Im\o\to 0$ this factor becomes real.
This establishes the hermiticity of $S_{ab}(\myvec{p},\o)$
provided that $\Re\o$ is not equal to $E_n(\myvec{p})$ for any
intermediate state, so that $S_{ab}(\myvec{p},\o)$ is well defined.
In particular, it follows that $\car(\myvec{p})$
is hermitian except at the degeneracy points.
As explained in Sec.~\ref{nogo}, the inverse $\Heff(\myvec{p})$
is well defined and hermitian everywhere in the Brillouin zone.

\subsection{\label{Rpsmooth} Smoothness}
We will consider a hamiltonian $H$ satisfying the following conditions:

\mynext
(1)
$H$ depends on fermion fields only.

\mynext
(2)
It is possible to express $H$ as a sum
$H = \sum_{\myvec{x}} \ch(\myvec{x})$, where the hamiltonian density
$\ch(\myvec{x})$ is a {\em local composite operator}.\\
We define a local composite operator $\cb(\myvec{x})$
by the following requirements.
First, $\cb(\myvec{x})$ is translationally covariant.  Second,
$\cb(\myvec{x})$ is the sum of a finite number of terms, where each term
is the product of a finite number of elementary (fermion) fields
with its own (finite) coupling constant.  Finally, the support
of $\cb(\myvec{x})$ is limited to lattice sites $\myvec{y}$
whose distance from $\myvec{x}$ is bounded.  The smallest
$0 \le R_\cb < \infty$ such that $\|\myvec{y}-\myvec{x}\|\le R_\cb$ for all
$\myvec{y}$ in the support of $\cb(\myvec{x})$
is the {\em range} of $\cb(\myvec{x})$.
We will denote the range $R_\ch$ of the hamiltonian density
as $R_0$ for short.%
\footnote{
  The norm $\|\myvec{z}\|$ can be for instance the usual $L_2$ norm
  $\left(\sum_{i=1}^d z_i^2\right)^{1/2}$
  or the taxi-driver's distance $\sum_{i=1}^d |z_i|$.
  We note that the hamiltonian must contain hopping terms,
  and thus $R_0\ge 1$.
}

For simplicity we will consider here two-point functions of
the elementary fermion fields, which in turn satisfy the
canonical anti-commutation relations. The generalization
to the case that some fermion fields are local composite operators
is straightforward, and is left for the reader.

We start with some preliminaries.  The canonical anti-commutation relations
imply that, as an operator acting on the Fock space, the norm
of an elementary fermion field is bounded by $\|\j_a(\myvec{x})\|=1$
(in lattice units).
It follows from our assumptions that the norm of
the hamiltonian density $\cn=\|\ch(\myvec{x})\|$ is finite, $0<\cn<\infty$.

We next turn to the time-dependent field,
\begin{equation}
\label{psit}
\j_a(\myvec{x},t) = e^{iHt} \j_a(\myvec{x}) e^{-iHt} \ .
\end{equation}
The norm of the time-dependent anti-commutator satisfies the trivial bound
\begin{equation}
\label{bound2}
\| \{ \j_a(\myvec{x},t), \j_b^\dagger(\myvec{0},0) \} \| \le 2 \ .
\end{equation}
Introducing multi-commutators of the hamiltonian $H$ with
an elementary fermion field
\begin{eqnarray}
\label{multic}
[[H,\j_a]]_0 &=& \j_a \ ,
\\[0ex]
[[H,\j_a]]_1 &=& [H,\j_a] \ ,
\nonumber\\[0ex]
[[H,\j_a]]_2 &=& [H,[[H,\j_a]]_1] \ = \ [H,[H,\j_a]] \ ,
\nonumber\\[0ex]
[[H,\j_a]]_n &=& [H,[[H,\j_a]]_{n-1}] \ ,
\nonumber
\end{eqnarray}
the time-dependent field can be expressed in terms of the
canonical field at $t=0$ as
\begin{equation}
\label{Taylort}
\j_a(\myvec{x},t)=\sum_{n=0}^\infty \frac{(it)^n}{n!} [[H,\j_a(\myvec{x})]]_n \ .
\end{equation}
This Taylor series controls the locality properties of correlation functions,
as we will see.

We will now prove that the retarded function $\hat{R}(\myvec{p},t)$
is infinitely differentiable for all $\myvec{p}$ in the Brillouin zone.
Provided  $\Im\o>0$, the same is true for $\tilde{R}(\myvec{p},\o)$,
which is also an analytic function of $\o$.  The advanced functions
$\hat{A}(\myvec{p},t)$ and $\tilde{A}(\myvec{p},\o)$ have similar
properties, except the support of $\tilde{A}(\myvec{p},\o)$
is the lower half-plane $\Im\o<0$.

We begin by introducing  the function $n(\myvec{x})$, defined as the
smallest integer such that $n(\myvec{x}) \ge \taxi{\myvec{x}}/R_0$.
This means that the term with $n=n(\myvec{x})$ in
the Taylor series~(\ref{Taylort}) is the first term where the support
of $[[H,\j_a(\myvec{x})]]_n$ may include the origin.
It follows that the first term in Eq.~(\ref{Taylort})
that contributes to the anti-commutator
$\{ \j_a(\myvec{x},t), \j_b^\dagger(\myvec{0},0) \}$
must have $n\ge n(\myvec{x})$.
Estimating the norm of the anti-commutator
using the norm of the $n= n(\myvec{x})$ term we arrive at
\begin{equation}
\label{nacomm}
\| \{ \j_a(\myvec{x},t), \j_b^\dagger(\myvec{0},0) \} \|
\sim \frac{(2\cn t)^{n(\myvec{x})}}{n(\myvec{x})!}
\sim \left(\frac{2eR_0\cn t}{\taxi{\myvec{x}}}\right)^{\frac{\taxi{\myvec{x}}}{R_0}}
\ .
\end{equation}
The rightmost expression becomes smaller than one for
\begin{equation}
\label{compsuppx}
\taxi{\myvec{x}} > 2eR_0 \cn t \ .
\end{equation}
For fixed $t$, the estimate~(\ref{nacomm})
vanishes faster than exponentially for $\myvec{x}\to\infty$.

We now use the bounds~(\ref{bound2}) and~(\ref{nacomm}) to establish the
analyticity properties of the Fourier transform $\hat{R}(\myvec{p},t)$.
In this subsection we consider the case that $\myvec{p}$ is real.
In App.~\ref{Rpwedge} we will extend the discussion to complex $\myvec{p}$.
In the ``cone''
\begin{equation}
\label{suppx}
\taxi{\myvec{x}} \le 2eR_0 \cn t \ ,
\end{equation}
the norm $\| \{ \j(\myvec{x},t), \j^\dagger(\myvec{0},0) \} \|$
can only be bounded by 2, a bound which is always valid.
For $\taxi{\myvec{x}} \gg 2e R_0\cn t$ we may use Eq.~(\ref{nacomm}),
and the norm of the anti-commutator vanishes faster than exponentially.
Since $\hat{R}(\myvec{p},t)$ involves a spatial sum, it follows that
this sum is dominated by the region defined by inequality~(\ref{suppx}).
This implies that the norm of $\hat{R}(\myvec{p},t)$ behaves like
\begin{equation}
\label{hatRnorm}
\sim C(2e R_0\cn t)^d \ ,
\end{equation}
where again $d$ is the number of spatial dimensions,
and $C$ is a geometrical factor.\footnote{
  In this appendix we use the notation $C$ for all the geometrical factors
  we encounter, but it should be noted that they are in general
  different from each other.
}
Similarly, the $k$-th derivative of $\hat{R}(\myvec{p},t)$ with respect
to $\myvec{p}$ behaves like $\sim C(2e R_0\cn t)^{d+k}$, since each
$\myvec{p}$-derivative adds one power of $\myvec{x}$.  It follows that
$\hat{R}(\myvec{p},t)$ and all its $\myvec{p}$-derivatives exist.
Thus, $\hat{R}(\myvec{p},t)$ is infinitely differentiable
for any $\myvec{p}$ in the Brillouin zone, for any fixed $t\ge 0$.

We next perform the time Fourier transform.  Allowing for $k$ derivatives
of $\hat{R}(\myvec{p},t)$ with respect to $\myvec{p}$, together with
$m$ additional $\o$-derivatives, the integrand in Eq.~(\ref{Retcxt}) behaves like
\begin{equation}
\label{tbound}
\sim C(2e R_0\cn t)^{d+k} t^m e^{-t\,\Im\o} \ .
\end{equation}
Hence, the $t$-integral converges for any $\Im\o>0$.\footnote{
  This is where the use of the retarded (or advanced)
  two-point function is crucial.
}
It thus follows that $\tilde{R}(\myvec{p},\o)$ is an analytic function
of $\o$ in the upper half plane, $\Im\o>0$, as well as an
infinitely differentiable function of $\myvec{p}$ everywhere in the
Brillouin zone.
The advanced function $\tilde{A}(\myvec{p},\o)$ has similar properties,
except that it is analytic in the lower half plane, $\Im\o<0$.
Finally, by the arguments of Sec.~\ref{nogo} it follows that
the common $\o\to 0$ boundary value $\car(\myvec{p})=\ca(\myvec{p})$
is infinitely differentiable with respect to $\myvec{p}$,
except at the degeneracy points.

\subsection{\label{Rpwedge} Analyticity}
The results of App.~\ref{Rpsmooth} are sufficient for the no-go theorem,
which only requires a continuous first derivative.  Nevertheless,
for completeness we will use here the edge-of-the-wedge theorem
to obtain the stronger result that $\car(\myvec{p})$ is an analytic function
of $\myvec{p}$ except at the degeneracy points.\footnote{%
  For the precise statement and a proof of the edge-of-the-wedge theorem,
  see for example Ref.~\cite{edgeproof}.
}

In order to apply the edge-of-the-wedge theorem we need to specify
the ``edge'' and the two ``wedges.''  We start with the edge $\ce$,
which will be an open subset of the real $(\myvec{p},\o)$ space,
where $\myvec{p}$ belongs to the Brillouin zone.
For each $\myvec{p}$, let $\Emin(\myvec{p})\ge 0$ be the infimum
(greatest lower bound) of the energy of the states
of the second-quantized hamiltonian with this total momentum.
We then define an interval $\O(\myvec{p})$ as the open subset
of the (real) $\o$ axis with $|\o| < \Emin(\myvec{p})$.
The edge $\ce$ is defined as the union
of $\O(\myvec{p})$ for all $\myvec{p}$ in the Brillouin zone.
Note that the degeneracy points are {\em not} included in $\ce$,
because if $\myvec{p}$ is a degeneracy point then $\Emin(\myvec{p})=0$
and thus $\O(\myvec{p}_c)$ is an empty set.
For any other $\myvec{p}$, the edge $\ce$ contains the open interval
$\O(\myvec{p})$ which, in particular, includes the point $\o=0$.
It follows that for real $\o\in \O(\myvec{p})$ the (common) limit
\begin{equation}
\label{limRA}
\lim_{\e\to 0} \tilde{R}(\myvec{p},\o+i\e) =
\lim_{\e\to 0} \tilde{A}(\myvec{p},\o-i\e)
\end{equation}
exists, and is continuous in $\ce$.

The two wedges are defined as $W^\pm=\ce\pm i\cu$, where $\cu$ is
the intersection of an open cone in
$\textbf{R}^{d+1}=(\Im\myvec{p},\Im\o)$ with a ball of radius $r>0$.
(An open cone $\cu$ is an open set such that if $u\in \cu$
then $su\in \cu$ for any $s>0$.) We next discuss the choice of $\cu$.
In App.~\ref{Rpsmooth} we already allowed $\o$ to have
an imaginary part, and now we seek a generalization to the case
that $\myvec{p}$ has an imaginary part as well.
First, using Eq.~(\ref{nacomm}), for $\taxi{\myvec{x}} \gg 2e R_0\cn t$
the summand in Eq.~(\ref{Retcx}) is bounded by
\begin{equation}
\label{boundImp}
\sim e^{\taxi{\Im\myvec{p}} \taxi{\myvec{x}} }
\left(\frac{2eR_0\cn t}{\taxi{\myvec{x}}}\right)^{\frac{\taxi{\myvec{x}}}{R_0}}
=
\left(\frac{2R_0\cn t\, e^{R_0\taxi{\Im\myvec{p}}+1}}{\taxi{\myvec{x}}}
  \right)^{\frac{\taxi{\myvec{x}}}{R_0}} \ .
\end{equation}
This still vanishes faster than exponentially
for asymptotically large $\taxi{\myvec{x}}$.
Hence $\hat{R}(\myvec{p},t)$ exists and is continuous as before,
and the same is true for all of its derivatives with respect to $\myvec{p}$,
where now $\myvec{p}$ can take complex values.  It follows that
$\hat{R}(\myvec{p},t)$ is an analytic function of $\myvec{p}$.

Estimating the norm of $\hat{R}(\myvec{p},t)$ requires more care.
We begin by noting that Eq.~(\ref{boundImp}) becomes
smaller than one when $\taxi{\myvec{x}}$ becomes larger than
$s_0(\Im\myvec{p}) t$, where (compare Eq.~(\ref{compsuppx}))
\begin{equation}
\label{suppxImp}
s_0(\Im\myvec{p}) = 2R_0 \cn e^{R_0\taxi{\Im\myvec{p}}+1} \ .
\end{equation}
It follows that we can neglect the region
$\taxi{\myvec{x}} > s_0(\Im\myvec{p}) t$.
We need to perform the $\myvec{x}$-summation over the complement region
$\taxi{\myvec{x}} \le s_0(\Im\myvec{p}) t$
and, as before, in that region we will use the
trivial bound~(\ref{bound2}) on the anti-commutator, which is applicable
everywhere.  The summand is then bounded by
\begin{equation}
\label{crudeb}
2 e^{\taxi{\Im\myvec{p}} \taxi{\myvec{x}} }
\le 2 e^{\taxi{\Im\myvec{p}} s_0(\Im\myvec{p}) t}
\ \le 2 e^{\taxi{\Im\myvec{p}} s_0^{\rm max} t} \ ,
\end{equation}
where $s_0^{\rm max}$ is the maximum of $s_0(\Im\myvec{p})$
for $\|\Im\myvec{p}\|\le r$.
Using the right-hand side as a uniform bound
for $\taxi{\myvec{x}} \le s_0(\Im\myvec{p}) t$ leads to
the following (over)-estimate\footnote{%
  The generalization to the case that derivatives with respect
  to $\myvec{p}$ and $\o$ are taken works as before
  (compare Eqs.~(\ref{hatRnorm}) and~(\ref{tbound})).}
of the norm of $\hat{R}(\myvec{p},t)$
\begin{equation}
\label{hatRImp}
\sim C (s_0(\Im\myvec{p}) t)^d e^{\taxi{\Im\myvec{p}} s_0^{\rm max} t} \ .
\end{equation}
In order for the $t$-integral to converge, we thus impose
\begin{equation}
\label{cone}
\taxi{\Im\myvec{p}}\, s_0^{\rm max} < \Im\o \ .
\end{equation}
This condition defines the open cone, and thus the open set $\cu$
as its intersection with the ball of radius $r$, for all $\Im\o>0$.
It then follows that $\tilde{R}(\myvec{p},\o)$ is an analytic function
of both $\myvec{p}$ and $\o$ in the wedge $\ce+i\cu$.
The same is true for $\tilde{A}(\myvec{p},\o)$ in the wedge $\ce-i\cu$.

Having defined an edge and two wedges that satisfy the assumptions
of the edge-of-the-wedge theorem, the theorem asserts
that there exists an open set $\cd\subset \textbf{C}^{d+1}$ such that
(a) $\cd$ contains the union of $W^+$, $W^-$ and $\ce$;
(b) there is an analytic function $F(\myvec{p},\o)$ defined on $\cd$
whose restriction to $W^+$ ($W^-$) is $\tilde{R}(\myvec{p},\o)$
($\tilde{A}(\myvec{p},\o)$), and whose restriction
to the edge $\ce$ is given by Eq.~(\ref{limRA}).
In particular, for $\o=0$ we have $F(\myvec{p},0)=\car(\myvec{p})$,
provided that $\myvec{p}$ is not a degeneracy point.\footnote{%
  On physical grounds we expect that for general real $\myvec{p}$,
  the analytic function $F(\myvec{p},\o)$ will have
  two cuts in the complex $\o$ plane along the real $\o$ axis:
  one starts at $\o = \Emin(\myvec{p})$ and goes to $+\infty$,
  and the other starts at $\o = -\Emin(\myvec{p})$ and goes to $-\infty$.
  At the degeneracy points, the end points of the two cuts meet at $\o=0$.}
This completes
the proof that $\car(\myvec{p})$ is an analytic function of $\myvec{p}$
everywhere in the Brillouin zone except at the degeneracy points.

\subsection{\label{secondary} Secondary singularities}
Here we explain the role of secondary singularities via
several examples involving four-fermion interactions.
We first consider an example of a self-energy correction
near a primary singularity, of the kind already discussed in Sec.~\ref{nogo}.
We then consider an example of a self-energy correction near
a secondary singularity, and discuss the similarities and differences
between the two cases.  As both examples are somewhat abstract, we
give a further example how they can be realized in the context of
weakly coupled theories.

To keep things as simple as possible
we will consider a Brillouin zone extending over the standard interval
$[0,2\p/a]$ for every momentum component.  In $d=1$, we will assume
that primary singularities exist at $p=0$ and $p=\pi/(2a)$.
In $d=3$ we will assume the same situation for $p_x$,
while the $p_y$ and $p_z$ components of the primary singularities
under consideration are always zero.
Below, we will focus on the massless states associated with
the primary singularity for which $p$ (or $p_x$) equals $\pi/(2a)$,
which we will eventually use in our example of a secondary singularity.

We begin with the self-energy correction for a massless fermion
at the primary singularity $p_c=\pi/(2a)$ for $d=1$,
or $\vec{p}_c=(\pi/(2a),0,0)$ for $d=3$.  To avoid cumbersome notation
we will mostly omit the vector symbol below, but the discussion applies
to both $d=1$ and $d=3$.  We assume that the collection of massless states
associated with primary singularities at $p_c$ is described by
a set of lattice interpolating fields $\j_i$, $i=1,\ldots,N$,
with conserved charges $Q_i(\j_j)=\d_{ij}$
under the corresponding U(1) symmetries.\footnote{%
  Handedness plays little role in the argument, and is therefore suppressed.
}
As explained in Sec.~\ref{nogo}, self-energy corrections
can be calculated using an EFT approach, and
the leading (momentum dependent) self-energy correction is a two-loop
diagram with two four-fermion vertices.
We will also make contact with the discussion
in Sec.~\ref{nogo2dim} by presenting examples in which the
four-fermion interactions are renormalizable in $d=1$.
For this to be the case, the four-fermion interactions have to exist already
at the lattice level without the need of point splitting (which turns
into additional derivatives in the continuum limit).
In the case at hand we assume $N=2$,
and then take the four-fermion interaction to be the local operator
$(\j_1^\dagger \j_1) (\j_2^\dagger \j_2)$.  The resulting self-energy
is given in Eq.~(\ref{relz1log}) for $d=1$ and Eq.~(\ref{relz3log}) for $d=3$.
This self-energy diagram corresponds to a process
in which an initial state of a single $\j$ particle with small $q=p-p_c$
splits into a virtual state consisting of two particles
and one anti-particle, which then recombine.
To produce the singularities in Eq.~(\ref{relzlog}),
the lattice momenta of the two particles have to be close to $p_c$,
whereas the lattice momentum of the anti-particle is close to $-p_c$.
This allows the process to occur while preserving
the lattice momentum.\footnote{%
  Since an anti-particle is associated with a hole in the Dirac sea,
  in the charge $Q=-1$ sector the primary singularity
  of $\car(\vec{p})$ is at $-\pi/(2a).$}

We next turn to an example of a self-energy diagram near
a secondary singularity.  In our example there is a secondary singularity
at $p_s=3\pi/(2a)$, associated with an intermediate state
of three $\j$ fermions.
Now the single-particle state on the external legs will in general be
a different fermion species, which we will denote as $\c$.
We again avoid the need of point splitting (or equivalently, derivatives)
by assuming $N=3$, and then taking the local four-fermion interaction to be
$(\c^\dagger \j_1\j_2\j_3 + \hc\!)$.
This interaction preserves in particular the U(1) symmetry
associated with a common rotation of all the $\j$ fermions,
with charge $Q=Q_1+Q_2+Q_3=3$, under which $Q(\c) = 3$.
The four-fermion interaction now enables the process
of a single $\c$ particle splitting (or decaying) into three $\j$ particles.
Since a single $\c$ intermediate state also contributes in the same channel,
by the definition of a secondary singularity it should be gapped near $p_s$,
in other words, its energy for $p=p_s$ should be $|E_0|>0$.
Let us introduce $q_i = p_i - p_c$ for the three $\j$ virtual particles,
as well as $q=q_1+q_2+q_3$ and $p=p_1+p_2+p_3$, so that $q = p-p_s$.
A straightforward diagrammatic calculation of the
self-energy correction for the $\c$ particle, coming
from the three massless $\j$ intermediate states,
gives rise for small $q$ to the schematic form (compare Eq.~(\ref{relzlog})),
\begin{subequations}
\label{2nd}
\begin{eqnarray}
\label{2nd1log}
E &=& E_0 \pm c_1 G^2 q (aq)^{2(n-d-1)} \log(q^2) + \cdots \ ,
\hspace{8.4ex} d=1 \ ,
\\
\label{2nd3log}
H_{2\times 2} &=& E_0 \pm c_3 G^2 \,
\vec\s\cdot\vec{q}\, (aq)^{2(n-d-1)} \log(q^2) + \cdots  \ ,
\qquad d=3 \ . \hspace{5ex}
\end{eqnarray}
\end{subequations}
where again $n$ is the mass dimension of the interaction, and
the ellipsis indicate subleading and/or analytic corrections.
As in Eq.~(\ref{relzlog}), $G$ is a dimensionless coupling constant.
Our discussion implies that
$\Heff$ has a finite value at the secondary singularity,
but it is not analytic there, in agreement with the general considerations
of Sec.~\ref{nogo}.  In the example we have considered here,
for $d=3$ the four-fermion interaction is irrelevant: one has $n-d-1=2$,
hence expression~(\ref{2nd3log}) has four continuous derivatives.
In contrast, for $d=1$ the four-fermion interaction is marginal: $n-d-1=0$.
As a result, expression~(\ref{2nd1log}) is continuous,
but does not have a continuous derivative at $p_s$, as discussed
in Sec.~\ref{nogo2dim}.

In summary, we see that the non-analytic behavior near
secondary singularities is similar to that near the primary ones.
The main difference is that at the primary singularities $\Heff$ has
zero eigenvalues, which are required to be relativistic,
whereas at the secondary singularities it does not.
In fact, at a generic point in the phase diagram
any secondary singularity is inherently non-relativistic,
because the gapped single-particle state in the relevant channel has
energy $E_0$ which is $O(1)$ in lattice units.  Nevertheless,
the behavior of $\Heff$ near the secondary singularities
must be understood, because the NN theorem requires a continuous
first derivative throughout the entire Brillouin zone.

Primary and secondary singularities occur in the two-point functions
of both weakly interacting and strongly interacting theories.
The latter case includes SMG models, which are difficult to study.
We can make the discussion
more concrete by demonstrating how the primary and secondary singularities
arise in the appropriate two-point functions of weakly coupled
lattice theories, which can be studied systematically
using perturbation theory. Of course, in our weakly coupled examples,
the massless fermions can always be combined into Dirac fermions,
and the theory is vectorlike; the goal is mainly
to illustrate the nature of primary and secondary singularities.

For the example of the primary singularity, all we need to do
is specify a bilinear hamiltonian for the $\j_i$ fields,
which, specializing to $d=1$, can be taken to be
\begin{equation}
\label{Hpsi}
H_i = \frac{\sqrt{2}}{a} \int_0^{2\p/a} \frac{dp}{2\p}\,
\j_i^\dagger(-p) [\cos(ap-\pi/4) – \cos(\pi/4)] \j_i(p)\ ,
\end{equation}
in momentum space.
It is easy to check that this hamiltonian supports
one RH and one LH massless states, at $p=0$ and $p=p_c=\p/(2a)$ respectively,
each leading to a primary singularity in the two-point function
$\langle\j_i\,\j_i^\dagger\rangle$.  Once the (by assumption, weak)
four-fermion interaction $(\j_1^\dagger \j_1) (\j_2^\dagger \j_2)$
is added, we recover the analytic structure discussed above near both of these
primary singularities.

For the example of the secondary singularity, we need a bilinear hamiltonian
for the additional $\c$ field, which we simply take to be
\begin{equation}
\label{Hchi}
H_\c = \frac{1}{a} \int_0^{2\p/a} \frac{dp}{2\p}\, \c^\dagger(-p) \sin(ap) \c(p)\ .
\end{equation}
When the interaction $(\c^\dagger \j_1\j_2\j_3 + \hc\!)$
is added, the $\c$ field couples to three-particle states
of the $\j$ fields, and some of these states will have vanishing energy
near $p_s=3\pi/(2a)$.  The single $\c$ state
with momentum $p \sim 3\pi/(2a)$ is gapped with energy
$|E_0|=a^{-1}+\co(G^2)$ for $p=p_s$, hence we will
indeed recover the secondary singularity discussed above
in the two-point function $\langle\c\,\c^\dagger\rangle$.

The presence of the renormalizable four-fermion interaction in $d=1$
implies that, in certain channels, the resulting $\Heff$
will not have a continuous first derivative (for primary singularities,
see Eq.~(\ref{relz1log}), for secondary singularities, see Eq.~(\ref{2nd1log}) above).
As a result, one of the main assumptions of the generalized no-go theorem
is not satisfied.  As we have explained in Sec.~\ref{nogo2dim},
under these circumstances one can tentatively still identify RH (LH)
massless states with a branch $E(p)$ of $\Heff$ that crosses zero
from negative to positive (positive to negative) energy.
The weakly coupled lattice hamiltonians presented here are examples where
this identification is valid.  For more subtle situations where
this is not (or not necessarily) the case, see Sec.~\ref{nogo2dim}.

Notice that we have assumed that the primary singularity is at
$\pi/(2a)$, which is equal to $2\p/a$ times $1/4$, a rational number.
It was shown in Refs.~\cite{NNYSPRL,NNYSlong} that this is a completely general
feature: every momentum component of a primary singularity must be equal
to $2\pi/a$ times a rational number. The same is then also true
for all the secondary singularities.  This implies that one can define
a reduced $d$-dimensional Brillouin zone
in which all the (finitely many) primary and
secondary singularities collapse to $\vec{k}=0$.
The momentum $\vec{k}$
in this reduced Brillouin zone can then be identified
with the physical momentum, which transforms homogeneously under
Lorentz transformations in the continuum limit.
For the full discussion, see Refs.~\cite{NNYSPRL,NNYSlong}.

\vspace{5ex}

\end{document}